\documentclass[11pt]{article}
\pdfoutput=1
\usepackage{jheppub}
\usepackage{graphicx}
\usepackage{amssymb}
\usepackage{amsmath}
\usepackage{amsfonts}
\usepackage{latexsym}
\usepackage{hyperref}
\usepackage{physics}
\usepackage{bbm}
\usepackage{empheq}
\usepackage{tikz}
\usepackage{cancel}
\usepackage{hhline}
\usepackage{caption}
\usepackage{soul}
\usepackage{float}
\usepackage{multirow}
\usepackage{hhline}
\usepackage{algorithm}
\usepackage{algpseudocode}
\usepackage{comment}

\usepackage[T1]{fontenc}

\usetikzlibrary{arrows.meta}
\usetikzlibrary{decorations.markings}
\tikzset{->-/.style={decoration={
			markings,
			mark=at position #1 with {\arrow{>}}},postaction={decorate}}}

\newcommand\mI{\mathcal{I}}

\newcommand{\KKeven}{\mathbb{K}^{\text{even}}}

\newcommand{\sfX}{\textsf{X}}

\newcommand{\sfM}{\textsf{M}}
\newcommand{\sfd}{\textsf{d}}

\makeatletter
\newcommand*{\rom}[1]{\expandafter\@slowromancap\romannumeral #1@}
\makeatother
\begin{document}

\begin{flushright}
\small
\texttt{HU-EP-25/15}
\end{flushright}
    
\title{Analytical and numerical routes to strong coupling\\ in $\mathcal{N}=2$ SCFTs
}
	
\author[a]{Pieter-Jan De Smet,}
\author[b]{Alessandro Pini,}
\author[c,d]{Paolo Vallarino\,}

\affiliation[a]{Institute for Theoretical Physics, KU Leuven\,,\\ Celestijnenlaan 200D, B-3001 Leuven, Belgium}
\affiliation[b]{Institut f{\"u}r Physik, Humboldt-Universit{\"a}t zu Berlin,\\
     IRIS Geb{\"a}ude, Zum Großen Windkanal 2, 12489 Berlin, Germany}
\affiliation[c]{ Universit\`a di Torino, Dipartimento di Fisica,\\
			Via P. Giuria 1, I-10125 Torino, Italy}
\affiliation[d]{I.N.F.N. - sezione di Torino,\\
			Via P. Giuria 1, I-10125 Torino, Italy }
\emailAdd{pieterjan.desmet@kuleuven.be}
\emailAdd{alessandro.pini@physik.hu-berlin.de}
\emailAdd{paolo.vallarino@unito.it}


\abstract{
We consider the $\mathcal{N}=2$ quiver gauge theory arising from a $\mathbb{Z}_M$ orbifold of $\mathcal{N}=4$ Super Yang-Mills theory. Over the years, exploiting supersymmetric localization, exact expressions for several observables have been derived in the planar limit of this theory. In particular, some of these can be expressed as Fredholm determinants of semi-infinite matrices and their strong coupling expansions in inverse powers of the 't Hooft coupling have been calculated  analytically to any desired order. On the other hand, there are also observables that cannot be rewritten in such a closed form, therefore extracting information at strong coupling  is more complicated and almost no results are known beyond the leading order.
In this work we focus on two observables of this type: the correlators of 
$n$ coincident Wilson loops and the integrated correlators of two Higgs branch operators in the presence of a Wilson line.
We introduce an analytic method to evaluate the first terms of their strong coupling expansions. We also outline a numerical algorithm that serves as an independent check of the analytical results and provides predictions in cases where analytical techniques are currently not known.
}

\maketitle \flushbottom

\section{Introduction}\label{sec:intro}
The study of superconformal field theories (SCFTs) has attracted significant attention in recent years. The large amount of symmetry in these models constrains their dynamics, enabling the use of different analytical techniques which have led to important progress in understanding the strong coupling behavior of such gauge theories. In particular, considerable effort has been devoted to the study of $\mathcal{N}=4$ SYM, the maximally supersymmetric theory in four dimensions, primarily through the use of integrability, the AdS/CFT correspondence, and supersymmetric localization. The latter method can also be applied in the context of theories with a lower amount of supersymmetry, such as $\mathcal{N}=2$ SCFTs, yielding, in the planar limit, many results for different observables that are exact, namely valid for any value of the 't Hooft coupling. These include the partition function of the theory \cite{Pestun:2007rz}, the vacuum expectation value of half-BPS Wilson loops \cite{Passerini:2011fe,Beccaria:2021vuc}, correlation functions between chiral/antichiral operators \cite{Gerchkovitz:2016gxx,Baggio:2016skg,Rodriguez-Gomez:2016ijh,Rodriguez-Gomez:2016cem,Billo:2017glv,Beccaria:2020hgy,Beccaria:2021hvt,Bobev:2022grf,Pini:2017ouj} and correlators between Wilson loops and chiral operators \cite{Billo:2018oog,Pini:2023svd,Sysoeva:2017fhr}.

A particularly interesting four-dimensional $\mathcal{N}=2$ SCFT is provided by the $M$-node circular quiver gauge theory obtained as a $\mathbb{Z}_M$ orbifold of $\mathcal{N}=4$ SYM. Part of the importance of this theory lies in the fact that it can be regarded as the starting point for the study of other $\mathcal{N}=2$ SCFTs, which can subsequently be obtained by performing further orientifold projections, see e.g.~\cite{Ennes:2000fu}. Moreover, recently techniques have been developed to systematically study the large $N$ limit of this quiver gauge theory. The main computational difficulty, compared to $\mathcal{N}=4$ SYM, is due to the fact that supersymmetric localization in $\mathcal{N}=2$ theories leads to a matrix model with a non-trivial potential, making the derivation of exact expressions very challenging. However, the quiver gauge theory at the orbifold fixed point, i.e. when all the gauge couplings associated with different nodes are equal, admits a gravity dual \cite{Kachru:1998ys,Gukov:1998kk}; in this specific configuration of the theory, as shown in \cite{Billo:2021rdb,Billo:2022fnb}, the partition function can be expressed in terms of a semi-infinite matrix, called $\textsf{X}$-matrix, which is given by a convolution of Bessel functions of the first kind. In turn, this enabled the exact evaluation of many observables in the large 
$N$ limit of the theory \cite{Mitev:2014yba,Mitev:2015oty,Pini:2017ouj,Galvagno:2020cgq,Billo:2021rdb,Billo:2022gmq,Billo:2022fnb,Beccaria:2022ypy,Billo:2022lrv,Rey:2010ry,Zarembo:2020tpf,Ouyang:2020hwd,Fiol:2020ojn,Galvagno:2021bbj,Preti:2022inu,Beccaria:2023kbl,Beccaria:2023qnu,Sobko:2025zci,Pini:2024uia}. Moreover, it was shown in \cite{Beccaria:2022ypy} that, in the planar limit, some observables, such as the free energy, the vacuum expectation value of the half-BPS Wilson loop and the two-point function between a chiral and an antichiral operator, admit exact expressions in terms of Fredholm determinants of the $\textsf{X}$-matrix\,\footnote{In \cite{Beccaria:2022ypy} the $\mathsf{X}$-matrix was related to a mathematical object known as Bessel operator.}. This turns out to be particularly useful at strong coupling, as the large 't Hooft coupling expansion of these quantities can be derived algorithmically to any desired order, thus providing information on the non-perturbative regime of the theory and serving as the starting point for a subsequent resurgence analysis, as recently carried out in \cite{Bajnok:2024epf,Bajnok:2024ymr}. Furthermore, these results offer predictions for holographic computations, beyond the supergravity approximation \cite{Skrzypek:2023fkr}. 
It is therefore natural to investigate whether the strong coupling expansion of other observables can also be performed by exploiting the asymptotic expansion of Fredholm determinants of Bessel operators. An important step in this direction was the recognition that, in the planar limit, the three-point function of chiral and antichiral operators can be factorized into a product of differential operators acting solely on two-point functions. This was first proven in \cite{Billo:2022lrv} for the two-node quiver gauge theory and was recently extended to the generic $M$-node quiver in \cite{Korchemsky:2025eyc,Ferrando:2025qkr}.

Nevertheless, there are also more complex observables for which exact expressions in the large $N$ limit can be derived, but which cannot be rewritten in terms of Fredholm determinants of Bessel operators. It is then important to develop alternative approaches to derive the corresponding strong coupling expansions. In this work, we aim to take a step in this direction by introducing both analytical and numerical methods to tackle this problem. Our analytical approach is based on the so-called ``method of differential equations'' developed in \cite{Belitsky:2020qrm,Belitsky:2020qir, Beccaria:2023kbl}, which was extended to the context of the $M>2$ node quiver gauge theory in \cite{Pini:2023lyo,Beccaria:2023qnu}. To the best of our knowledge, the strong coupling expansion of observables that are {\it not} Fredholm determinants of a Bessel operator have been computed only at leading order, with the only exception being the non-planar corrections to the free energy of the quiver gauge theory \cite{Beccaria:2023qnu}.
In this work, we present a method to analytically compute the first subleading corrections. Moreover, we also outline a numerical approach based on solving an  integral equation. This method should be regarded as complementary to the analytical one, as it allows for independent verification of analytical predictions.  Furthermore, it provides results in cases where computations become too involved to be performed analytically or where no analytical tools are available. In recent years, the construction of Padé approximants has been the most commonly employed numerical technique in this context, as it provides valuable information at very large values of the coupling and often serves as input for resurgence analyses \cite{Beccaria:2021vuc,Costin:2020hwg,Costin:2020pcj}. The main advantage of the method introduced in this article is that it generally yields more accurate predictions for the coefficients of the strong coupling expansion of a given observable compared to those obtained via Padé approximants, also requiring lower computational cost.

To illustrate these methods, we consider the $\mathcal{N}=2$ quiver gauge theory, focusing on two specific observables. The first one is the correlator of $n$-coincident Wilson loops \cite{Pini:2023lyo}, for which both analytical and numerical approaches can be applied. Secondly, to provide an example where the use of the numerical method becomes indispensable, we also consider an integrated correlator between a Wilson line and two moment map operators \cite{Pufu:2023vwo,Billo:2023ncz} in the specific case of two-node quiver.  
However, it is important to remark that, as argued in the conclusions, our techniques can be applied to a broader set of $\mathcal{N}=2$ theories and observables.

The rest of this paper is organized as follows. In Section \ref{sec:ZMquiver}, we review the main properties of the $4d$ $\mathcal{N}=2$ quiver gauge theory, together with the two observables relevant to our analysis, and explain how their evaluations in the large $N$ limit can be efficiently carried out using an interacting matrix model. In Section \ref{sec:NumMethod}, we introduce the numerical method employed to obtain predictions at strong coupling. In Section \ref{sec:WL}, we examine the correlator of $n$-coincident Wilson loops and compute the first terms of its strong coupling expansion. In Section \ref{sec:intcorr}, we analyze the integrated correlator mentioned above, for which numerical methods are essential, and derive the leading terms of its large $\lambda$ expansion. Finally, we present our conclusions in Section \ref{sec:conclusions}.

\section{The \texorpdfstring{$\mathbb{Z}_M$}{} quiver gauge theory and the observables}\label{sec:ZMquiver}
\begin{figure}[ht]
	\center{\includegraphics[scale=0.9]{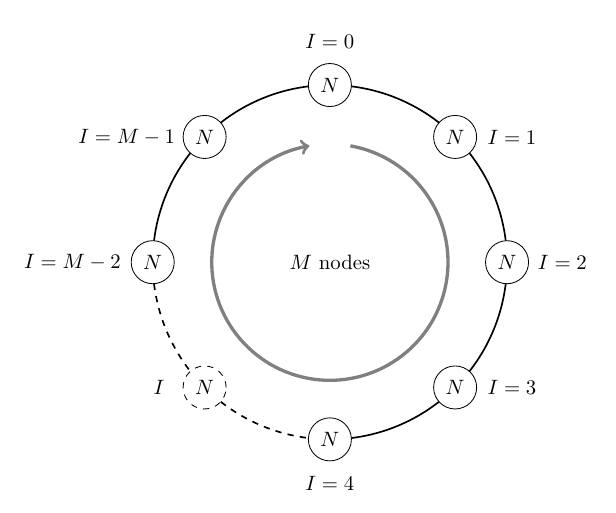}
		\caption{A graphical representation of the $\mathbb{Z}_M$ quiver theory with $M$ nodes.
		\label{fig:1_quiver}} } 
\end{figure}

We construct the $\mathbb{Z}_M$ quiver gauge theory through an orbifold projection from the $\mathcal{N}=4$ SYM theory with gauge group $SU(M N)$. We can engineer this set-up  by considering a stack of $M N$ D3-branes in Type IIB superstring theory placed on a $\mathbb{C}^2/\mathbb{Z}_M$ orbifold singularity. Breaking this configuration into $M$ stacks of $N$ fractional D3-branes located at the orbifold fixed-point, we obtain the quiver theory with $M$ nodes depicted in Fig.~\ref{fig:1_quiver} (for details see \cite{Douglas:1996sw,Kachru:1998ys,Skrzypek:2023fkr}). The quiver gauge theory thus obtained preserves $\mathcal{N}=2$ supersymmetry. Each node, labeled by $I=0,\dots,M-1$, is associated with an $SU(N)$ gauge group, whereas the links connecting them represent bifundamental hypermultiplets\,\footnote{All over the paper the index $I$ is taken modulo $M$, that is to say $I \sim I + M$.}. It is straightforward to check that such matter content guarantees a vanishing  $\beta$-function in each node and then conformal symmetry at the quantum level.
In particular, we focus on the quiver with equal gauge couplings in all the nodes, which has a well established holographic dual, given by Type IIB superstring theory on the $AdS_5\times S^5/\mathbb{Z}_M$ space \cite{Kachru:1998ys,Gukov:1998kk}.

The main gauge invariant operator that we consider is the half-BPS circular Wilson loop of unit radius in the fundamental representation of the gauge group of the $I$-th node of the quiver, which is defined as \cite{Maldacena:1998im,Semenoff:2001xp} 
\begin{align}
\label{WilsonI} 
W^{(I)} \equiv \frac{1}{N}\,  \textrm{tr} \, \mathcal{P} \, \textrm{exp} \,  \Bigg\{ g \oint_{\mathcal{C}} \, d\tau \left[ i\,A^{I}_{\mu}\dot{x}^{\mu}(\tau) + \frac{1}{2}\left(\varphi^{I}(x)+\overline{\varphi}^{I}(x)\right)\right] \Bigg\} \,,
\end{align}
where $\mathcal{C}$ is the circle, $\mathcal{P}$ denotes the path-ordering, while $A_{\mu}^{I}$ and $\varphi^{I}$ are the gauge field and the complex scalar belonging to the $I$-th vector multiplet and $g$ the coupling constant. 
In the quiver gauge theory, following the analysis carried out in \cite{Rey:2010ry,Galvagno:2021bbj,Pini:2023lyo}, it is convenient to introduce specific combinations of these Wilson loop operators:
\begin{subequations}
\begin{align}
& W_0 \equiv \frac{1}{\sqrt{M}}\left(W^{(0)}+W^{(1)}+\, \cdots \, + W^{(M-1)}\right)\,, \label{Wuntwisted} \\
& W_{\alpha} \equiv \frac{1}{\sqrt{M}}\sum_{I=0}^{M-1}\rho^{I\alpha}\,W^{(I)}\,, \label{Wtwisted}
\end{align}
\label{Wall}
\end{subequations}
where $\alpha=1,\dots, M-1$ and $\rho$ denotes the $M$-th root of unity, namely
\begin{align}
\label{rho}
\rho \equiv \text{e}^{2\pi i/M}\, .
\end{align}
We refer to the operator \eqref{Wuntwisted} as untwisted Wilson loop and to the operators  \eqref{Wtwisted} as twisted Wilson loops. Of course the two definitions \eqref{Wall} can be combined into
\begin{align}
\label{Wfull}
W_{\alpha} \equiv \frac{1}{\sqrt{M}}\sum_{I=0}^{M-1}\rho^{I\alpha}\,W^{(I)}\, \ \, ,
\end{align}
where $\alpha=0,1,\dots,M-1$.

We now introduce the two distinct observables involving the untwisted and twisted Wilson loops that we aim to analyze.
\subsubsection*{Correlators of $\mathbf{n}$ coincident Wilson loops}
The first quantity we examine
is the planar term of correlators among multiple coincident Wilson loops, namely
\begin{align}
\label{Wcorrelator}
\big \langle W_{\alpha_1}\,W_{\alpha_2}\dots W_{\alpha_n} \big \rangle \,.
\end{align}
We observe that these $n$-point correlators vanish unless
\begin{align}
\label{ZMchargetozero}
\sum_{i=1}^{n} \alpha_i = 0 \ \text{mod}  \ M\,,
\end{align}
as the $\mathbb{Z}_M$ symmetry, which  rotates the nodes of the quiver, requires  that the total ``twist charge'' of any observable must be zero modulo $M$. Therefore, henceforth we assume that $\alpha_n = -\alpha_1 -\alpha_2  \dots- \alpha_{n-1}$. Furthermore, as already observed, for instance in \cite{Pini:2023lyo}, it is important to recall that the $n$-point function of coincident untwisted Wilson loops is planar equivalent to its counterpart in $\mathcal{N}=4$ SYM theory; additionally, mixed correlators involving $n$ untwisted and $m$ twisted Wilson loops, at planar order in the large $N$ expansion, factorize into the product of the $n$-point function of untwisted Wilson loops and the $m$-point function of twisted ones. Hence, in this work, we only focus on correlators with coincident twisted Wilson loops.

\subsubsection*{An integrated correlator}
The second observable we consider is an integrated correlator between two moment map operators of conformal dimension two in the presence of a Wilson line. This is a special class of correlation functions that has gained increasing attention in recent years. It was originally introduced in the context of $\mathcal{N}=4$ SYM theory \cite{Pufu:2023vwo,Billo:2023ncz,Dempsey:2024vkf,Billo:2024kri,Dorigoni:2024vrb,Dorigoni:2024csx} and later extended also to $\mathcal{N} = 2$ theories \cite{Dempsey:2024vkf,Pini:2024zwi,DeLillo:2025hal}.

Importantly, it can still be computed using supersymmetric localization. Specifically, this requires considering the mass-deformed theory, where the hypermultiplets acquire a mass, and then taking two mass-derivatives of the logarithm of the Wilson line vacuum expectation value, namely
\begin{align}
\label{intcorrW}
\mathcal{W} \equiv \partial_m^2\log \big\langle W \big\rangle\big\vert_{m=0} = \int\!d^4x_1\,d^4x_2\,\mu(x_1,x_2) \big\langle \mathcal{O}_2(x_1)\,\mathcal{O}_2(x_2) \big\rangle_W \, , 
\end{align}
where, as shown in \cite{Billo:2023ncz,Dempsey:2024vkf,Billo:2024kri}, the integration measure $\mu(x_1, x_2)$ is uniquely determined by superconformal symmetry. In this formula $\mathcal{O}_2(x_i)$ represents a dimension-two moment map operator\,\footnote{These operators correspond to the top component of the short multiplet $\hat{\mathcal{B}}_1$ within the $\mathfrak{su}(2,2|2)$ superconformal algebra.} and $\langle \mathcal{O}_2(x_1)\,\mathcal{O}_2(x_2) \big\rangle_W$ denotes the 2-point function between these operators in the presence of the Wilson line. After integration, the information about the original space-time structure is lost, leaving us only with a non-trivial function of the coupling. Nevertheless, an important aspect of the quantity defined in \eqref{intcorrW} is that it can be used as a constraint in numerical boostrap computations (see e.g. \cite{Chester:2022sqb}) or for holographic calculations on AdS \cite{Alday:2024srr}.

\subsection{The matrix model}
\label{subsec:MatrixModel}
In a seminal paper \cite{Pestun:2007rz} Pestun showed that the partition function of a four-dimensional $\mathcal{N}=2$ supersymmetric theory placed on a sphere $S^4$ can be computed exactly by exploiting supersymmetric localization. Using this result, the partition function $\mathcal{Z}$ of the $\mathcal{N}=2$ quiver gauge theory can be written as finite dimensional integrals over a collection of $M$ matrices $a_I = a_{I}^bT_b$ taking values in the $\mathfrak{su}(N)_I$ algebra\,\footnote{$T_b$ are the generators of the
$SU(N)$ Lie algebra in the fundamental representation, normalized such that 
\begin{align}
\text{tr}\,T_bT_c = \frac{1}{2}\delta_{b,c}\,,\qquad b,c=1,\cdots, N^2-1 \,.
\end{align}} 
\begin{align}
\label{Zpartition}
\mathcal{Z} = \int \left(\prod_{I=0}^{M-1} \, da_I \right) \, \text{e}^{-\text{tr}\,a_I^2}\, \big\vert\mathcal{Z}_{\text{1-loop}}\,\mathcal{Z}_{\text{inst}}\big\vert^2 \, ,
\end{align}
where $\mathcal{Z}_{\text{1-loop}}$ encodes the contributions coming from the fluctuations around the localization locus, while $\mathcal{Z}_{\text{inst}}$ is the instanton partition function. The latter can be neglected since we focus on the large $N$ limit of the theory and, henceforth, we set $\mathcal{Z}_{\text{inst}}=1$. We employ the so called ``full-Lie algebra approach'' \cite{Billo:2017glv,Fiol:2018yuc}, in such a way that integrations are performed over all the matrix elements $a_I^b$. Moreover, the integration measure is defined as
\begin{align}
da_I = \prod_{b=1}^{N^2-1}\frac{da^{b}_I}{\sqrt{2\pi}}\, 
\end{align}
ensuring that the Gaussian integration over each $a_I$ equals 1. Importantly, 
we can recast the $\mathcal{Z}_{\text{1-loop}}$ in terms of an interaction action, namely
\begin{align}
\big\vert\mathcal{Z}_{\text{1-loop}}\big\vert^2 = \text{e}^{-S_{\mathrm{int}}} \,,
\end{align}
where \cite{Billo:2021rdb}
\begin{align}
\label{Sint}
S_{\mathrm{int}} = \sum_{I=0}^{M-1}\left[\,
\sum_{m=2}^{+\infty}\sum_{k=2}^{2m}(-1)^{m+k}\Big(\frac{\lambda}{8\pi^2N}\Big)^{m}\,\binom{2m}{k}\,
\frac{\zeta_{2m-1}}{2m}\,\big(\textrm{tr} \, a_I^{2m-k}-\textrm{tr}\,a_{I+1}^{2m-k}\big) \big(\textrm{tr}\, a_I^k - \textrm{tr} \, a_{I+1}^k\big) \, \right],
\end{align}
with $\zeta_{2m-1}$ denoting the odd Riemann $\zeta$-values. In this way the vacuum expectation value of a generic observable $f(a)$ becomes
\begin{align}
\big\langle f(a) \big\rangle = \frac{\big\langle f(a) \, \text{e}^{-S_{\text{int}}} \big\rangle_0}{\big\langle \text{e}^{-S_{\text{int}}} \big\rangle_0} \,,
\end{align}
where $\big\langle \cdot \big\rangle_0$ stands for the v.e.v. in the free matrix model. The expectation values in the Gaussian matrix model can then be evaluated by exploiting extremely efficient recursion relations (see e.g. \cite{Billo:2017glv}) satisfied by
\begin{align}
t_{n_1,n_2,\ldots,n_q} = \big\langle \text{tr}a^{n_1}\text{tr}a^{n_2}\cdots\text{tr}a^{n_q} \big \rangle\,  .    
\end{align}
In the matrix model of the $\mathbb{Z}_M$ quiver gauge theory, it is convenient to introduce 
twisted and untwisted linear combinations of the traces of the matrices $a_I$ which are useful to represent the operators of the gauge theory, such as the Wilson loop \eqref{WilsonI}. Hence, following \cite{Billo:2021rdb,Billo:2022fnb}, we define the operators 
\begin{align}
\label{Aoperators}
A_{\alpha,k}=\frac{1}{\sqrt{M}}\sum_{I=0}^{M-1}\rho^{\alpha I}\,\textrm{tr}\,a_I^k
\end{align}
with the understanding that $A_{\alpha,k}^\dagger=A_{M-\alpha,k}$. In particular $\alpha =0$ corresponds to the untwisted combination, while  $\alpha\neq 0$ labels the twisted ones.
Furthermore, we introduce the vevless basis of the $A_{\alpha,k}$'s operators, i.e.
\begin{align}
\label{Ahatoperators}
\hat{A}_{\alpha,k}\equiv A_{\alpha,k}-\big\langle A_{\alpha,k} \big\rangle \,,
\end{align}
where, from \eqref{Aoperators}, it follows that
\begin{align}
\label{Atwisted}
\hat{A}_{\alpha,k}=A_{\alpha,k}\ \ \ \ \text{for}\ \ \alpha\neq 0\,.
\end{align}
This basis is particularly useful to study the large $N$ limit of correlators in the matrix model. 
In particular, in \cite{Billo:2021rdb,Billo:2022fnb} it was found that the leading behavior of 2- and 3-point functions is
\begin{subequations}
\begin{align}
\label{Wick2A}
& \big\langle \hat{A}_{\alpha,k}\,\hat{A}_{\beta,\ell}^{\dagger} \big\rangle_0  \, \propto \, N^{\frac{k+\ell}{2}}\delta_{\alpha,M-\beta} \ \ \text{with} \ \ k+\ell \ \  \text{even}\,, \\[0.5em]
& \big\langle\hat{A}_{\alpha,k}\,\hat{A}_{\beta,\ell}\,\hat{A}_{\gamma,q}^{\dagger} \big\rangle_0 \, \propto \, N^{\frac{k+\ell+q}{2}-1}\delta_{\alpha+\beta,M-\gamma}\, , \ \text{with} \ k+\ell+q \ \ \text{even}\,.
\label{Wick3A}
\end{align}
\label{WickA}%
\end{subequations}
Moreover, it was shown that the leading order of the large $N$ expansion of a generic higher point correlator can be factorized à la Wick:  into products of 2-point functions when the number of $\hat{A}_{\alpha,k}$
operators is even, or into products of one 3-point function and 2-point functions when the number is odd. So far, we have considered only the Gaussian theory. However, the particular expression of the interaction action \eqref{Sint} allows this factorization property to be extended to the interacting theory as well. For the purposes that follow, it is useful to consider the following change of basis \cite{Beccaria:2020hgy,Billo:2021rdb}
\begin{align}
\label{fromAtoP}
\hat{A}_{\alpha,k} = \left(\frac{N}{2}\right)^{\frac{k}{2}}\sum_{i=0}^{\lfloor\frac{k-1}{2}\rfloor}\sqrt{k-2i}\left(\begin{array}{c}
     k  \\
     i 
\end{array}\right)\mathcal{P}_{\alpha,k-2i} \,.
\end{align}
Indeed, rewriting the interaction action \eqref{Sint} in this basis, one obtains
\begin{align}
\label{SintPbasis}
S_{\mathrm{int}} = -\frac{1}{2}\sum_{\alpha=0}^{M-1}\sum_{k,\ell=2}^\infty s_{\alpha}\,\mathcal{P}_{\alpha,k}^{\dagger}\,\textsf{X}_{k,\ell}\,\mathcal{P}_{\alpha,\ell} \,,
\end{align}
where
\begin{align}
\label{salfa}
s_{\alpha} \equiv \sin\left(\frac{\pi\alpha}{M}\right)^2 \,,
\end{align}
and $\textsf{X}$ is a semi-infinite matrix, whose entries with opposite parity vanish
\begin{align}
\label{XEvenOdd}
\textsf{X}_{2k,2\ell+1} = 0 \,,
\end{align}
while its non-trivial elements read
\begin{align}
\label{Xmatrix}
\textsf{X}_{k,\ell} = -8\,(-1)^{\frac{k+\ell+2k\ell}{2}}\,\sqrt{k\,\ell}\,\int_0^{\infty} \frac{dt}{t} \frac{\text{e}^{t}}{(\text{e}^{t}-1)^2}\,J_k\left(\frac{t\sqrt{\lambda}}{2\pi}\right)J_{\ell}\left(\frac{t\sqrt{\lambda}}{2\pi}\right)
\end{align}
with $k,\ell \geq 2$. For our purposes, it is helpful to define the following sub-matrices
\begin{align}
\sfX^{\text{even}}_{k,\ell} &\equiv \mathsf{X}_{2k,2\ell} = -\int_0^{\infty}\!\!\!\! dt\ U_k^{\text{even}}(t)\ U_\ell^{\text{even}}(t)\,, \label{eq:Xeven}\\
\sfX^{\text{odd}}_{k,\ell}&\equiv \mathsf{X}_{2k+1,2\ell+1} = -\int_0^{\infty}\!\!\!\! dt\ U_k^{\text{odd}}(t)\ U_\ell^{\text{odd}}(t)\,, \label{eq:Xodd}\\
\intertext{with} 
U_k^{\text{even}}(t) &= (-1)^k  \frac{2\,\sqrt{k}}{\sqrt{t}\,\sinh (t/2)}   \,J_{2 k} \left( \frac{t \sqrt{\lambda} } {2 \pi} \right)\,,\label{eq:Vk}\\
U_k^{\text{odd}}(t) &= (-1)^k \frac{\sqrt{2(2k+1)}}{\sqrt{t}\,\sinh (t/2)} \, J_{2 k+1} \left( \frac{t \sqrt{\lambda} } {2 \pi} \right)\,.\label{eq:Wk}
\end{align}
The main importance of the expression \eqref{Xmatrix} relies on the fact that the coupling dependence, which in \eqref{Sint} appeared as a perturbative expansion, is now resummed in terms of Bessel functions of the first kind.
Consequently, performing
matrix model computations using the interaction action \eqref{SintPbasis} permits to obtain exact expressions for the observables in terms of the $\mathsf{X}$-matrix. For instance, the partition function of the quiver gauge theory becomes \cite{Billo:2021rdb}\,\footnote{Henceforth, we use the symbol "$\simeq$" to denote the leading term in the large $N$ expansion.}
\begin{equation}
\mathcal{Z} \, \simeq \, \text{det}^{-1} \left(\mathbf{1}-s_\alpha\,\mathsf{X} \right)\,,
\end{equation}
and a similar analysis has also been conducted for correlation functions of $\mathcal{P}_{\alpha,k}$'s operators in \cite{Billo:2021rdb,Billo:2022fnb}. Specifically, for 2-point functions, it was found that
\begin{align}
\label{2pointP}
& \big\langle \mathcal{P}_{\alpha,k}\,\mathcal{P}_{\beta,\ell}^{\dagger} \big\rangle \simeq \delta_{\alpha,\beta}\,\textsf{D}_{k,\ell}^{(\alpha)} \,,
\end{align}
where
\begin{align}
\label{Dalfa}
\textsf{D}_{k,\ell}^{(\alpha)} = \left(\frac{1}{1-s_{\alpha}\,\textsf{X}}\right)_{k,\ell}\,.
\end{align}
Also here we  define
\begin{align}
\mathsf{D}_{k,\ell}^{(\alpha)\,\text{even}}\equiv \mathsf{D}_{2k,2\ell}^{(\alpha)} \,,\qquad\qquad \mathsf{D}_{k,\ell}^{(\alpha)\,\text{odd}} \equiv \mathsf{D}_{2k+1,2\ell+1}^{(\alpha)}\,.
\label{Deando}
\end{align}
For the 3-point functions it holds
\begin{align}
\label{3pointP}
& \langle\, \mathcal{P}_{\alpha,k}\,\mathcal{P}_{\beta,\ell}\,\mathcal{P}_{\gamma,q}^{\dagger}\, \rangle \simeq \delta_{\alpha+\beta,\gamma}\,\frac{\textsf{d}_{k}^{(\alpha)}\textsf{d}_{\ell}^{(\beta)}\textsf{d}_{q}^{(\gamma)}}{\sqrt{M}\,N} \qquad \text{with}\ \ k+\ell+q\ \ \text{even} \, ,
\end{align}
where
\begin{align}
\label{dalfa}  
\textsf{d}^{(\alpha)}_{k} = \sum_{\ell=2}^{\infty}\textsf{D}_{k,\ell}^{(\alpha)}\,\sqrt{\ell} \, .
\end{align}
Higher point functions, at the leading order in the large $N$ expansion, can still be computed using Wick's theorem. This is because the $\mathcal{P}_{\alpha,k}$ operators are linear combination of the $\hat{A}_{\alpha,k}$ operators, and thus they inherit the properties \eqref{WickA}.

These results allow for the exact computation of several quantities in the interacting matrix model of the quiver gauge theory. From here on, we will focus solely on the two observables of interest for this paper, considering them separately.

\subsubsection{Correlators of $\mathbf{n}$ coincident Wilson loops}

The matrix model representation of the Wilson loops \eqref{Wfull}
is \cite{Pestun:2007rz}
\begin{align}
\label{defWL}
W_{\alpha} = \frac{1}{\sqrt{M}\,N}\sum_{I=0}^{M-1}\rho^{\alpha I}\textrm{tr}\,\textrm{exp}\biggl[\sqrt{\frac{\lambda}{2N}}a_I \biggr]=\frac{1}{N}\sum_{k=0}^{\infty}\frac{1}{k!}\left(\frac{\lambda}{2N}\right)^{\frac{k}{2}}A_{\alpha,k} \, .
\end{align}
From the expression above, it follows immediately that the evaluation of correlators of $n$ coincident Wilson loops is recast into the computation of correlation functions among $A_{\alpha,k}$ operators, which can then be rewritten in terms of the $\mathcal{P}_{\alpha,k}$ operators using the change of basis \eqref{fromAtoP}. In particular, the planar term of this observable can be evaluated by exploiting the matrix model techniques discussed in Section \ref{subsec:MatrixModel}.  This analysis was performed in \cite{Pini:2023lyo}, where exact expressions for the 2-point and 3-point correlators have been obtained. We find it useful to recall these results here, as they serve as starting point for the strong coupling analysis that will be performed in Section \ref{sec:WL}.  
Specifically, it was found that the ratio between the correlator of 2 coincident twisted Wilson loops and the 2-point connected correlator $W^{(2)}_{conn}= \langle W_0W_0 \rangle_0 -\langle W _0 \rangle_0^2$ in $\mathcal{N}=4$ SYM 
is given by
\begin{align}
\label{eq:ratioWW}
\frac{\big\langle W_{\alpha}W_{\alpha}^{\dagger} \big\rangle}{ W_{conn}^{(2)}(\lambda)} \equiv 1 +\Delta w^{(\alpha)}(M,\lambda) \,,
\end{align}
with
\begin{align}
\label{eq:w}
\Delta w^{(\alpha)} (M,\lambda) \simeq  \frac{2}{\sqrt{\lambda}} \sum_{k=2}^{\infty}\sum_{\ell=2}^{\infty}\frac{I_k(\sqrt{\lambda})\,I_{\ell}(\sqrt{\lambda})}{I_{1}(\sqrt{\lambda})\,I_{2}(\sqrt{\lambda})}\,\sqrt{k\,\ell}\, (\textsf{D}_{k,\ell}^{(\alpha)}-\delta_{k,\ell})\,, 
\end{align}
where $I_n(\sqrt{\lambda})$ are modified Bessel functions of the first kind. 
The ratio between the 3-point correlator among twisted Wilson loops and the connected 3-point correlator\,\footnote{See the expression (5.12) of \cite{Pini:2023lyo} for its definition.} $W_{conn}^{(3)}(\lambda)$ in  $\mathcal{N}=4$ SYM is given by
\begin{align}
\label{ratioWWW}
\frac{\big\langle W_\alpha\,W_\beta\,W_{\alpha+\beta}^{\dagger} \big\rangle}{\sqrt{M}\,W_{conn}^{(3)}(\lambda)} \equiv 1+\Delta w^{(\alpha,\beta)}(M,\lambda)  \,, 
\end{align}
whose exact expression in the planar limit reads
\begin{align}
1+\Delta w^{(\alpha,\beta)} \, \simeq \, & \frac{8}{\lambda^{3/2}}\left(\frac{I_1(\sqrt{\lambda})^2}{I_1(\sqrt{\lambda})^2+3I_2(\sqrt{\lambda})^2}\right)
\left[\, \prod_{p=1}^{3}\left( \mathcal{S}_{\text{even}}^{\alpha_p}+ \frac{\sqrt{\lambda}}{2} \right) \right. \nonumber \\
 &+ \left.\sum_{\sigma \in \mathcal{Q}_3}
\left(\mathcal{S}_{\text{even}}^{\alpha_{\sigma(1)}}+ \frac{\sqrt{\lambda}}{2}\right)
\left(\mathcal{S}_{\text{odd}}^{\alpha_{\sigma(2)}}+ \frac{\sqrt{\lambda}}{2}\frac{I_2(\sqrt{\lambda})}{I_1(\sqrt{\lambda})}\right)
\left(\mathcal{S}_{\text{odd}}^{\alpha_{\sigma(3)}}+ \frac{\sqrt{\lambda}}{2}\frac{I_2(\sqrt{\lambda})}{I_1(\sqrt{\lambda})}\right) 
\right]\, ,
\label{ratioWaWbWc}
\end{align}
where $\alpha_1 \equiv \alpha\,, \alpha_2 \equiv \beta\,, \alpha_3 \equiv M-\alpha-\beta$. Moreover, the quantities $\mathcal{S}_{\text{even}}^{\alpha} $  and  $\mathcal{S}_{\text{odd}}^{\alpha} $ are given by
\begin{subequations}
\begin{align}
\mathcal{S}_{\text{even}}^{\alpha} 
&=  \sum_{k,\ell=1}^{\infty}\frac{I_{2k}(\sqrt{\lambda})}{I_1(\sqrt{\lambda})}\sqrt{(2k)(2\ell)}\left(\textsf{D}_{2k,2\ell}^{(\alpha)}-\delta_{2k,2\ell}\right)\,, \label{eq:Seven}\\
\mathcal{S}_{\text{odd}}^{\alpha} 
&=  \sum_{k,\ell=1}^{\infty}\frac{I_{2k+1}(\sqrt{\lambda})}{I_1(\sqrt{\lambda})}\sqrt{(2k+1)(2\ell+1)}\left(\textsf{D}_{2k+1,2\ell+1}^{(\alpha)}-\delta_{2k+1,2\ell+1}\right)\label{eq:Sodd}\, ,
\end{align}
\label{SevenandSodd}%
\end{subequations}
and the set of permutations is defined as
\begin{align}
\label{setQ3}
\mathcal{Q}_3 = \{\,(1,2,3),\,(3,1,2),\,(2,3,1)\, \}\,.
\end{align}
Finally, an exact expression for the generic correlator of $n$ coincident Wilson loops, valid in the planar limit, has also been derived starting from \eqref{eq:ratioWW} and \eqref{ratioWWW}, and exploiting the Wick factorization properties of the $\mathcal{P}_{\alpha,k}$ operators. Due to its complexity, we do not report it here and instead refer the reader to \cite{Pini:2023lyo} for a detailed presentation. However, in Section \ref{sec:WL} we discuss its strong coupling expansion.

\subsubsection{The integrated correlator}
With the primary aim to simplify the discussion and establish a connection with the results of \cite{Pini:2024zwi}, where exact expressions for the planar and next-to-planar terms of the integrated correlator \eqref{intcorrW} have been derived in the context of the 2-node quiver, we also choose to restrict our analysis to this specific theory. Therefore, we consider the mass-deformed partition function of the $\mathbb{Z}_2$ quiver gauge theory, which is obtained by giving masses $m_1$ and $m_2$ to the two bifundamental hypermultiplets. Following \cite{Pini:2024uia}, we then expand it up to order $O(m^2)$, obtaining
\begin{align}
\mathcal{Z}(m_1,m_2) = \int\! da_0 \int\! da_1\, \, \text{e}^{-(\text{tr}\,a_0^2+\text{tr}\,a_1^2)}\,\text{e}^{-S_{\text{int}}+(m_1^2+m_2^2)\,\mathcal{M} \, +\,  O(m^4)}\, ,  
\label{ZmExpansion}
\end{align}
where $S_{\text{int}}$ is given in \eqref{SintPbasis} and
\begin{align}
\mathcal{M} =-\sum_{n=1}^{\infty}\sum_{\ell=0}^{2n}(-1)^{n+\ell}\,\frac{(2n+1)!\,\zeta_{2n+1}}{(2n-\ell)!\,\ell!}\left(\frac{\lambda}{8\pi^2N}\right)^n\text{tr}\,a_0^\ell\,\text{tr}\,a_1^{2n-\ell} \, .
\label{MZ2} 
\end{align}
In principle, one could consider an integrated correlator involving both an untwisted and a twisted Wilson loop. However, it follows from the definition \eqref{Wtwisted} that the latter has a vanishing vacuum expectation value and we therefore no longer consider it. On the other hand, in the case of an untwisted Wilson loop $W_0$, the v.e.v. $\mathcal{W}$ of the corresponding integrated correlator reads \cite{Pini:2024zwi} 
\begin{align}
\mathcal{W} = 2\,\frac{\big\langle W_0\,\mathcal{M} \big\rangle-\big\langle W_0 \big\rangle \big\langle \mathcal{M} \big\rangle}{\big\langle W_0 \big\rangle} \,.
\label{intcorrWexact}
\end{align}
We focus on its large $N$ expansion, namely
\begin{align}
\mathcal{W} = \mathcal{W}^{(L)} + \frac{\mathcal{W}^{(NL)}}{N^2} + O\left(\frac{1}{N^4}\right)\,  .
\label{LargeNW}    
\end{align}
In particular the planar coefficient $\mathcal{W}^{(L)}$ coincides with  the $\mathcal{N}=4$ result derived in \cite{Billo:2023ncz,Billo:2024kri,Pufu:2023vwo} and it reads
\begin{align}
\label{Wleading}
\frac{2\pi\sqrt{\lambda}}{I_1(\sqrt{\lambda})}\int_0^{\infty}\frac{dt}{t}\frac{(t/2)^2}{\sinh(t/2)^2}\frac{1}{4\pi^2+t^2}J_1\left(\frac{t\sqrt{\lambda}}{2\pi}\right)\left[2\pi\,I_0(\sqrt{\lambda})J_1\left(\frac{t\sqrt{\lambda}}{2\pi}\right)-tI_1(\sqrt{\lambda})J_0\left(\frac{t\sqrt{\lambda}}{2\pi}\right)\right] \,.
\end{align}
We therefore concentrate only on $\mathcal{W}^{(NL)}$, for which an exact expression can also be determined. In particular, the numerator of \eqref{intcorrWexact} admits the large $N$ expansion
\begin{align}
\big\langle W_0\,\mathcal{M} \big\rangle-\big\langle W_0 \big\rangle \big\langle \mathcal{M} \big\rangle \equiv \big\langle W_0\,\mathcal{M} \big\rangle_{c} \, \simeq \, \big\langle W_0\,\mathcal{M} \big\rangle_{c}^{(L)}\,+\,\frac{\big\langle W_0\,\mathcal{M} \big\rangle_{c}^{(NL)}}{N^2}\,+\,O\left(N^{-4} \right) \,,
\end{align}
where the explicit expressions for $\big\langle W_0\,\mathcal{M} \big\rangle_{c}^{(L)}$ and
$\big\langle W_0\,\mathcal{M} \big\rangle_{c}^{(NL)}$ have been found in \cite{Pini:2024zwi}. Moreover, the denominator, given by the v.e.v of the untwisted Wilson loop, admits the expansion \cite{Erickson:2000af,Rey:2010ry} 
\begin{align}
\big\langle W_0 \big\rangle \simeq W^{(L)}\,+\,\frac{W^{(NL)}}{N^2}\,+\,O\left(N^{-4} \right)
\end{align}
with
\begin{subequations}
\begin{align}
& W^{(L)}=\frac{2\sqrt{2}I_1(\sqrt{\lambda})}{\sqrt{\lambda}}\,, \label{WL}\\
& W^{(NL)} = \frac{\sqrt{2}}{48}\left( \lambda\,I_0(\sqrt{\lambda})-14\sqrt{\lambda}\,I_1(\sqrt{\lambda})\right)-\frac{\lambda^{3/2}\partial_\lambda\mathcal{F}}{2\sqrt{2}}I_1(\sqrt{\lambda})  \,,
\label{WNL}
\end{align}
\label{WLandWNL}%
\end{subequations}
where $\mathcal{F}$ is the free energy of the massless theory. 
Starting from these expressions the first terms of the strong coupling expansion of $\mathcal{W}^{(NL)}$ will be computed in Section \ref{sec:intcorr}.

\section{Numerical method}\label{sec:NumMethod}

In this section we outline the numerical method employed to perform the strong coupling expansions of the observables studied in this work. It is worth recalling that the most challenging aspect of this analysis lies in evaluating the inverse of the $\textsf{X}$-matrix, which appears in its resolvent \eqref{Dalfa}. 
Indeed, for all the observables we consider, it is necessary to compute expressions of the form
\begin{equation}\label{eq:sumaDb}
\sum_{k,\ell=1}^{\infty} a_k\, \mathsf{D}_{k,\ell}^{(\alpha)}\,  b_\ell\, , 
\end{equation}
where $a_k$ and $b_k$ are non-trivial functions of the 't Hooft coupling. Hence, our goal is now to present a numerical algorithm that allows us to evaluate the expression \eqref{eq:sumaDb}. For the sake of simplicity, we can just focus on the even components,  $\mathsf{D}_{k,\ell}^{(\alpha)\,\text{even}}$, as the odd ones can be treated in a completely analogous way. 
As shown in \cite{Bobev:2022grf}, this quantity can be expressed as 
\begin{align}
\textsf{D}^{(\alpha)\,\text{even}}_{k,\ell} = \delta_{k,\ell} - \int_0^{\infty} U_k^{\text{even}}(t)Z^{(\alpha)}_{\ell}\left(t\right) \ , \label{eq1}
\end{align}
where $Z_{\ell}^{(\alpha)}$ is the solution of the integral equation 
\begin{align}
Z_{\ell}^{(\alpha)}(t) + s_{\alpha}\int_0^{\infty}\!dt'\, \KKeven(t,t')\, Z_{\ell}^{(\alpha)}(t') = s_{\alpha}\,U_\ell^{\text{even}}(t)
\label{eq2}
\end{align}
with $U_k^{\text{even}}(t)$ defined in \eqref{eq:Vk} and
\begin{align}
\KKeven(t,t') = \sum_{k=1}^{\infty}U_k^{\text{even}}(t)U_k^{\text{even}}(t')\,.
\label{KandU}
\end{align}
Multiplying both sides of \eqref{eq1} by $a_k$ and $b_{\ell}$, and of \eqref{eq2} by $b_{\ell}$ alone, and then performing the sums over the indices, we obtain
\begin{equation}
\sum_{k,\ell=1}^{\infty} a_k\, \textsf{D}^{{(\alpha)\,\text{even}}}_{k,\ell}\,  b_\ell 
 = \sum_{k=1}^{\infty}a_k\,b_k-\int_0^{\infty}\!\!\!\! dt\, A(t)\, Z^{(\alpha)}(t)\, ,
 \label{eq3}
\end{equation}
where $Z^{(\alpha)}(t)$ is the solution of the integral equation
\begin{equation}
Z^{(\alpha)}(t) + s_{\alpha} \int_0^{\infty}\!\!\!\! dt'\,\KKeven(t,t')\, Z^{(\alpha)}(t')\,= \,s_{\alpha} B(t)
\label{eqZodd}
\end{equation}
with 
\begin{equation}\label{eq:KAB}
Z^{(\alpha)}(t)=\sum_{k=1}^{\infty}b_{k}\,Z_{k}^{(\alpha)}(t)\, , \quad 
A(t) =\sum_{k=1}^{\infty} a_k\, U^{\text{even}}_k(t)\quad\text{and}\quad B(t) =\sum_{k=1}^{\infty} b_k\, U^{\text{even}}_k(t)\,.
\end{equation}

A particularly efficient way to solve the integral equations \eqref{eq3}-\eqref{eqZodd} is the  Nyrstr\"om method, which approximates integrals with finite sums. Specifically, if we consider a function $f(x)$ defined on $[0,\infty [$  that decays sufficiently fast at infinity, we can replace the infinite interval  $[0, \infty[$ with a finite interval $[0, L]$ and approximate the integral as follows 
\begin{equation}\label{eq:inttosum}
\int_0^{\infty}\!\!\! dx\ f(x) \approx \sum_{i=1}^m w_i f(x_i) \, ,
\end{equation}
where $x_i \in [0,L]$ for $i=1, \ldots,m$, and $w_i$ are weights that depend on the specific approximation method. In our case, we choose to employ Fej\'er quadrature, whose points for the interval $[-1,1]$ are given by
\begin{equation}\label{eq:Fejertheta}
x_k = \cos \theta_k,\quad\text{with}\quad 	\theta_k = (2 k -1) \frac{\pi}{2 m}, \quad k = 1 , 2, \ldots, m\,,
\end{equation}
and the corresponding weights $w_k$ are (see for example \cite{waldvogel2006fast})
\begin{equation}\label{eq:Fejerweights}
	w_k = \frac{2}{m} \left[ 1 - 2 \sum_{r =1}^{ \frac{m-1}{2} } \frac{\cos(2 r \theta_k)}{4 r^2 -1} \right] \,.
\end{equation}
To apply the Fej\'er quadrature to integrals over the interval $[0,L]$,
the quadrature points and weights can be derived from \eqref{eq:Fejertheta} and \eqref{eq:Fejerweights} by performing a linear change of variables\,\footnote{ The Fej\'er quadrature points and weights are not directly available in {\tt Mathematica}. However, there is code available at the Wolfram Function Repository that can be used for their calculation \cite{WolframFunctionRepository}.}. Moreover, the cut-off $L$ and the number of discretization points 
 $m$ are chosen sufficiently large to ensure the accuracy of the approximation in \eqref{eq:inttosum}. Following this procedure, we arrive at the following algorithm.

\begin{algorithm}[H]\label{alg}
	\caption{Numerical calculation of $\sum_{k,\ell=1}^{\infty} a_k\, \textsf{D}^{{(\alpha)\,\text{even}}}_{k,\ell}\,  b_\ell$}
	\begin{algorithmic}
\State \textbf{Input:} Cutoff $L$, number of discretization points $m$
		\State \textbf{Step 1:} Set $t_i$ and $w_i$ equal to the Fej\'er points and weights on the interval $[ 0, L]$,  with $i = 1, \ldots, m$ 
		\State \textbf{Step 2:} Calculate the vector $B_i = \sqrt{w_i} B(t_i)$
\State \textbf{Step 3:} Calculate the $m \times m$ matrix $\KKeven_{ij} = \sqrt{w_i}\ \KKeven(t_i,t_j)\sqrt{w_j}$
\State \textbf{Step 4:} Solve the linear system $ Z^{(\alpha)}_i + s_{\alpha}\sum_j \KKeven_{ij} Z^{(\alpha)}_j =s_{\alpha}\, B_i$ for the vector $Z^{(\alpha)}_i$
		\State \textbf{Output:} $\sum_k a_k b_k - \sum_i \sqrt{w_i} A(t_i) Z^{(\alpha)}_i $
	\end{algorithmic}
\end{algorithm}
We stress that, by following the same procedure, an analogous algorithm can be derived for quantities involving the odd resolvent $\mathsf{D}^{(\alpha)\,\text{odd}}_{k,\ell}$.

\section{Correlators of \texorpdfstring{$n$}{} coincident twisted Wilson loops}\label{sec:WL}
In this section we study the strong coupling behavior of the correlators of $n$ coincident twisted Wilson loops in the $\mathbb{Z}_M$ quiver gauge theory. As reviewed in Section \ref{sec:ZMquiver}, the exact expression in the planar limit for this observable was found in \cite{Pini:2023lyo} with localization, where also the leading term at strong coupling was computed analytically.
Here we extend this analysis up to 3 subleading orders in the strong coupling expansion. 
We begin with  the simplest case, i.e. the correlator with 2 Wilson loops in Subsection \ref{subsec:2pt}, then we examine the case with 3 twisted Wilson loops in Subsection \ref{subsec:3pt}, and we derive the strong coupling expansion of the most general correlator of $n$ coincident twisted Wilson loops in Subsection~\ref{subsec:npt}. For both the 2-point and 3-point correlators, we first perform the computation analytically and subsequently apply the numerical method described in Section \ref{sec:NumMethod}, to validate our results and to generate new predictions.

\subsection{The 2-point function \texorpdfstring{$\langle W_{\alpha} W_{\alpha}^{\dagger} \rangle$}{}} \label{subsec:2pt}
We calculate the large $\lambda$ expansion of the 2-point function of twisted Wilson loops, whose planar exact expression is reported in~\eqref{eq:w}. The final result of our analysis is
\begin{equation}\label{eq:DeltawLargelambda}
1 + \Delta w^{(\alpha)}(M,\lambda) \underset{\lambda \rightarrow \infty}{\sim} \kappa_0\left(1+  \kappa_1\frac{1}{\sqrt{\lambda}} + \kappa_2 \frac{1}{\lambda}+ \kappa_3 \frac{1}{\lambda^{3/2}} + O\left(\lambda^{-2}\right)\right)
\end{equation}
with 
\begin{subequations}
\begin{align}
\kappa_0 &= \frac{1}{s_{\alpha}} \left( \frac{\mI_0(s_{\alpha})}{2}\right)^2 \,,
\label{c0} \\
\kappa_1 & = 2 \,, \\
\kappa_2 & = 3- \frac{\pi\ \mI_1(s_{\alpha})}{2}\,, \label{c2}\\
\kappa_3 &=\frac{15}{4} -\frac{3}{2} \pi \mI_1 (s_{\alpha}) -\pi^2 \mI_1 (s_{\alpha})^2 \,, \label{c3}
\end{align}
\end{subequations}
where the functions $\mI_n$ are defined as\,\footnote{The integral in Eq.~\eqref{eq:defCurlyI2308.03848} only converges for $n < 1/2$. For $n \ge 1/2$, it is understood via analytic continuation. }~\cite{Beccaria:2023kbl} 
\begin{equation}\label{eq:defCurlyI2308.03848}
\mI_n(s_{\alpha}) = \int_0^{+\infty} \frac{dz}{\pi} z^{1 - 2n} \partial_z \log\left(1 + s_\alpha\,\sinh \left( \frac{z}{2} \right)^{-2} \right).
\end{equation}
For $n=0,1$ this integral takes the following values~\cite{Beccaria:2022ypy}
\begin{align}
\mI_0(s_{\alpha}) &= -\frac{\alpha}{M}\left( 1 - \frac{\alpha}{M} \right)2\pi \\
\intertext{and}
\mI_1(s_{\alpha}) &= -\frac{1}{2 \pi}\left[ \psi\left( \frac{\alpha}{M} \right)+\psi\left(1- \frac{\alpha}{M} \right)  -2 \psi(1)\right]\, ,
\label{I1}
\end{align}
where $\psi(x)$ is the digamma function. We compute the coefficients \eqref{c0}-\eqref{c2} both analytically and numerically, while we compute \eqref{c3} only numerically.

\subsubsection{Analytical calculation}
\label{subsec:AnalyticSP}
To determine analytically the strong coupling limit of \eqref{eq:w}, we first decompose it into its even and odd components, obtaining 
\begin{equation}
\Delta w^{(\alpha)}(M,\lambda) = \Delta w^{(\alpha)\,\text{even}}(M,\lambda) + \Delta w^{(\alpha)\,\text{odd}}(M,\lambda)\,,
\label{spliteando}
\end{equation}
with 
\begin{align}
\Delta w^{(\alpha)\,\text{even}}(M,\lambda) &= \frac{2}{\sqrt{\lambda}} 
\sum_{k,\ell=1}^{\infty} \frac{I_{2k}(\sqrt{\lambda}) I_{2\ell}(\sqrt{\lambda} )}{I_1(\sqrt{\lambda}) I_2(\sqrt{\lambda} )}\sqrt{2 k} \sqrt{2 \ell}
\left(\mathsf{D}_{k,\ell}^{(\alpha)\,\text{even}}-\delta_{k,\ell} \right)\label{eq:weven}\\
\intertext{and}
\Delta w^{(\alpha)\,\text{odd}}(M,\lambda) &= \frac{2}{\sqrt{\lambda}} 
\sum_{k,\ell=1}^{\infty} 
\frac{I_{2k+1}(\sqrt{\lambda}) I_{2\ell+1}(\sqrt{\lambda} )}{I_1(\sqrt{\lambda}) I_2(\sqrt{\lambda} )}\sqrt{2 k+1} \sqrt{2 \ell+1}
\left(\mathsf{D}_{k,\ell}^{(\alpha)\,\text{odd}}-\delta_{k,\ell} \right)\,. \label{eq:wodd}
\end{align}
Then, we rewrite each of the two terms in the r.h.s. of \eqref{spliteando} as 
\begin{align}
\Delta w^{(\alpha)}\underset{\lambda \rightarrow \infty}{\sim} \frac{2}{\sqrt{\lambda}}\sum_{P=0}^{\infty}\frac{\mathcal{S}^{(P)}}{\lambda^{P/2}}     
\end{align}
where the coefficients of the expansion read
\begin{align}
\mathcal{S}^{(P)} =\sum_{L+J=P}S^{(L,J)}_{\text{odd}}+S^{(L,J)}_{\text{even}}
\label{DefinitionSp}
\end{align}
with
\begin{subequations}
\begin{align}
& S_{\text{odd}}^{(L,J)} = \sum_{k=1}^{\infty}\sum_{\ell=1}^{\infty}\sqrt{2k+1}\,Q_{2L}^{(1)\text{odd}}(k)\,\sqrt{2\ell+1}\,Q_{2J}^{(2)\text{odd}}(\ell)\,\big\langle \psi_{2\ell+1}\Big|\frac{s_{\alpha}\textsf{X}}{1-s_{\alpha}\textsf{X}}\Big|\psi_{2k+1} \big\rangle\,  , \label{Sodd}\\
& S_{\text{even}}^{(L,J)} = \sum_{k=1}^{\infty}\sum_{\ell=1}^{\infty}\sqrt{2k}\,Q_{2L}^{(1)\text{even}}(k)\,\sqrt{2\ell}\,Q_{2J}^{(2)\text{even}}(\ell)\,\big\langle \psi_{2\ell}\Big|\frac{s_{\alpha}\textsf{X}}{1-s_{\alpha}\textsf{X}}\Big|\psi_{2k} \big\rangle\,  .
\label{Seven}
\end{align}
\end{subequations}
In the previous expressions $Q_{2L}^{(j)\text{odd}}(k)$, $Q_{2L}^{(j)\text{even}}(k)$ are polynomials in $k$ obtained by taking the ratio of Bessel functions for large values of their arguments, namely
\begin{align}
\frac{I_{2k+1}(\sqrt{\lambda})}{I_j(\sqrt{\lambda})} \equiv \sum_{s=0}^{\infty}\frac{Q_{2s}^{(j)\text{odd}}(k)}{\lambda^{s/2}}\, , \qquad  \frac{I_{2k}(\sqrt{\lambda})}{I_j(\sqrt{\lambda})} \equiv \sum_{s=0}^{\infty}\frac{Q_{2s}^{(j)\text{even}}(k)}{\lambda^{s/2}}\, .
\end{align}
Furthermore, we  introduced the functions
\begin{align}
\psi_k(x) = (-1)^{\frac{k}{2}(k-1)}\sqrt{k}\frac{J_k(\sqrt{x})}{\sqrt{x}}\, ,  
\end{align}
which, in turn, allow us to express the non-trivial elements of the $\textsf{X}$-matrix as
\begin{align}
\mathsf{X}_{k,\ell} = \big\langle \psi_k| \mathsf{X}|\psi_{\ell} \big\rangle  \,  .  
\end{align}
The analysis performed in \cite{Pini:2023lyo} shows that the expression \eqref{DefinitionSp} can be expressed as a linear combination of the following coefficients
\begin{align}
w_{n,m}^{(\ell)}(s_{\alpha}) = \big\langle (x\partial_{x})^{n}\phi^{(\ell)}(x) \Big| \frac{s_{\alpha}\mathsf{X}}{1-s_{\alpha}\mathsf{X}}\Big| (x\partial_{x})^m\phi^{(\ell)}(x) \big\rangle\, ,    
\label{wnmCoefficients}
\end{align}
with $n,m \geq 0$ and where $\phi^{(\ell)}(x) \equiv J_{\ell}(\sqrt{x})$ with $\ell=1,2$. In particular, it turns out to be very useful to organize the expression of $\mathcal{S}^{(P)}$ in terms of the coefficients $w_{n,m}^{(\ell)}$ as
\begin{align}
\mathcal{S}^{(P)} = \mathcal{S}_{0}^{(P)} + \mathcal{S}_{1}^{(P)} + \mathcal{S}_{2}^{(P)} + \mathcal{S}_3^{(P)} + \cdots \, ,
\label{SpStrong}
\end{align}
where $S_{j}^{(P)}$ encodes the contribution from the coefficients $w^{(\ell)}_{r,s}$ satisfying $r+s=P-j$. Exploiting the techniques of Appendix B of \cite{Pini:2023lyo}, we determine the first three terms, namely
\begin{subequations}
\begin{align}
& \mathcal{S}_0^{(P)} = (-2)^{P-2}\,\sum_{n+m=P} \left[w_{n,m}^{(1)} + w_{n,m}^{(2)} \right]\,  , \\
& \mathcal{S}_{1}^{(P)} = (-2)^{P-3}\sum_{n+m=P}\left[-(1+n^2+m^2)w_{n,m}^{(1)}+(5-n^2-m^2)w_{n,m}^{(2)}\right]\, , \\
& \mathcal{S}_2^{(P)} = (-2)^{P-5}\sum_{n+m=P} \left[f^{(1)}_{n,m}\,w_{n,m}^{(1)}+ f^{(2)}_{n,m}\,w_{n,m}^{(2)}\right]\,  ,
\end{align}
\label{SSPexpansions}
\end{subequations}
where 
\begin{subequations}
\begin{align}
& f^{(1)}_{n,m} = \frac{1}{3} \left(-3 m^4-8 m^3-6 m^2 n^2-12 m^2+20 m-3
   n^4-8 n^3-12 n^2+20 n+15\right)\,  , \\
& f^{(2)}_{n,m} = \frac{1}{3} \left(-3 m^4-8 m^3-6 m^2 n^2+24 m^2+56 m-3
   n^4-8 n^3+24 n^2+56 n+51\right)\,  .
\end{align}
\end{subequations}
At this point we recall that, in the strong coupling limit, the coefficients $w_{n,m}^{(\ell)}$ admit the following expansion \cite{Beccaria:2023kbl,Pini:2023lyo}
\begin{align}
w_{n,m}^{(\ell)}= \omega_{n,m}^{(\ell,0)}\,g^{n+m+1} + \omega_{n,m}^{(\ell,1)}\,g^{n+m} + \omega_{n,m}^{(\ell,2)}\,g^{n+m-1} + \cdots\,  ,
\label{wStrongCoupling}
\end{align}
where
\begin{align}
g \equiv \frac{\sqrt{\lambda}}{4\pi} \,  .
\label{gCoupling}
\end{align}
It then becomes evident that expression \eqref{SpStrong} is particularly useful, as one can show that the leading term in the large $\lambda$ expansion of $\Delta w^{(\alpha)}$ is due only to $\mathcal{S}_0^{(P)}$. Similarly, the next-to-leading term receives contributions only from $\mathcal{S}_0^{(P)}$ and $\mathcal{S}_1^{(P)}$, while the next-to-next-to-leading term is entirely determined by  $\mathcal{S}_0^{(P)}$, $\mathcal{S}_1^{(P)}$ and $\mathcal{S}_2^{(P)}$, and so on. In particular, the leading coefficient \eqref{c0} was computed in \cite{Pini:2023lyo}. Using both \eqref{SSPexpansions} and \eqref{wStrongCoupling}, we find  that the next-to-leading term is given by
\begin{align}
\frac{2}{\sqrt{\lambda}}\sum_{P=0}^{\infty}\,\sum_{n+m=P}\frac{1}{(-2\pi)^n(-2\pi)^m}\left[\frac{(n^2+m^2-2)\,\omega_{n,m}^{(0)}}{16\pi}+\frac{1}{4}\sum_{\ell=1}^{2}\omega^{(\ell,1)}_{n,m}\right]\,  ,
\end{align}
where $\omega^{(0)}_{n,m} \equiv \omega_{n,m}^{(\ell,0)}$. Then, using the generating functions $G^{(0)}(s_{\alpha},x,y)$ and $G^{(1)}(\ell,s_{\alpha},x,y)$ defined in \eqref{GeneratingFunctions}, the expression above can be rewritten as
\begin{align}
\frac{1}{\sqrt{\lambda}}\Big( \frac{1}{8\pi}\left( (x\partial_x)^2 + (y\partial_y)^2 -2 \right)G^{(0)}(s_{\alpha},x,y) + \frac{1}{2}\sum_{\ell=1}^{2}G^{(1)}(\ell,s_{\alpha},x,y)\Big)\Big|_{x=y=-2\pi}\, ,
\label{c1Analytic}
\end{align}
where, after taking the derivatives, we set $x=y=-2\pi$. Finally, using the expressions collected in Appendix~\ref{app:GeneratingFunctions} and the identity (B.49) of \cite{Pini:2023lyo}, the next-to-leading term simplifies to
\begin{align}
\frac{2}{s_{\alpha} \sqrt{\lambda}} \left( \frac{\mathcal{I}_0(s_{\alpha})}{2} \right)^2\,,
\end{align}
which, in turn, implies that $\kappa_1=2$.

In a completely similar manner, exploiting the expression \eqref{SSPexpansions}, one can show that the next-to-next-to-leading term  in the strong coupling expansion of $\Delta w^{(\alpha)}$ is obtained by considering
\begin{align}
& 2\pi\sum_{\ell=1}^{2}G^{(2)}(\ell,s_{\alpha},x,y) + \frac{1}{4}\left((x\partial_x)^2+(y\partial_y)^2\right)\sum_{\ell=1}^{2}G^{(1)}(\ell,s_{\alpha},x,y) \nonumber \\
& + \frac{1}{4}(G^{(1)}(1,s_{\alpha},x,y)-5\,G^{(1)}(2,s_{\alpha},x,y))-\frac{1}{64\pi}\left(22 - \frac{76}{3}(x\partial_x+y\partial_y) + \right. \nonumber \\
&  \left. 4((x\partial_x)^2+(y\partial_y)^2) + \frac{16}{3}((x\partial_x)^3+(y\partial_y)^3) -4(x\partial_{x})^2(y\partial_y)^2-2((x\partial_x)^4+(y\partial_y)^4) \right)G^{(0)}(s_{\alpha},x,y)\, ,
\label{c2Analytic}
\end{align}
where, after taking the derivatives, we set $x=y=-2\pi$. Finally, by using the expressions collected in Appendix \ref{app:GeneratingFunctions}, and after a lengthy but straightforward computation, we obtain
\begin{align}
\frac{1}{s_{\alpha}}\left(\frac{\mathcal{I}_0(s_{\alpha})}{2}\right)^2\left(3-\frac{\pi\,\mathcal{I}_1(s_{\alpha})}{2}\right)\, .    
\end{align}
It is interesting to note that in both \eqref{c1Analytic} and \eqref{c2Analytic}, all derivative contributions completely cancel out after setting $x=y=-2\pi$. This suggests that a more direct procedure for deriving the result might exist. Moreover, in principle, there are no theoretical obstructions in deriving $\kappa_3$ using the same analytical method. However, such computation is highly involved. For this reason, in the following section, we show how the above results can also be recovered numerically and further extended to provide a prediction for the coefficient $\kappa_3$.

\subsubsection{Numerical calculation}\label{sec:230803848NumMethod}
Now we show how the numerical algorithm explained in Section \ref{sec:NumMethod} can be applied to compute the quantities \eqref{eq:weven} and \eqref{eq:wodd}. In particular, it suffices to present the analysis for the even component, as the discussion proceeds identically for the odd component \eqref{eq:wodd}. In order to employ this numerical method, we thus express $\Delta w^{(\alpha)\,\text{even}}$ as 
\begin{equation}
\Delta w^{(\alpha)\,\text{even}} = - \frac{2}{\sqrt{\lambda} }  \frac{1}{ I_1(\sqrt{\lambda}) I_2(\sqrt{\lambda})}\int_0^{\infty}\!\! dt\ A(t) \, Z^{(\alpha)}(t)\, ,
\label{numweven}
\end{equation}
where $Z^{(\alpha)}(t)$ is the solution of the integral equation 
\begin{equation}
Z^{(\alpha)}(t) + s_{\alpha} \int_0^{\infty}\!\!\!\! du\,  \KKeven(t,u)\, Z^{(\alpha)}(u)= s_{\alpha}\, B(t)\,
\label{numweven2}
\end{equation}
with $\KKeven$ defined in \eqref{KandU}.
To obtain the functions $A(t)$ and $B(t)$ we use $  a_k\,=\,b_k\,=\,\sqrt{2k}\,I_{2k}(\sqrt{\lambda})\,$. Plugging these expressions  in \eqref{eq:KAB}  along with the definition of $U_k^{\text{even}}$ \eqref{eq:Vk}, we get
\begin{equation}\label{closedformAandB}
A(t) \,=\, B(t) \,= \, \frac{\sqrt{2\,t}\,\pi}{\left( 4\pi^2+t^2 \right)\sinh (t/2)}\left(\frac{t\,\sqrt{\lambda}}{2\pi}\,I_1(\sqrt{\lambda})\,J_0\left(\frac{t\,\sqrt{\lambda}}{2\pi}\right) - \sqrt{\lambda}\,I_0(\sqrt{\lambda})\,J_1\left(\frac{t\,\sqrt{\lambda}}{2\pi}\right) \right) \, ,
\end{equation}
where we used the summation formulas of Bessel functions that can be found, for instance, in Appendix A of \cite{Pini:2023lyo}. Therefore,  we are now able to apply the numerical algorithm to solve the integral equations \eqref{numweven}-\eqref{numweven2}.

Before exploiting the algorithm for strong coupling calculations, we aim to furnish further evidence of its validity by testing it against perturbative computations. The expansions for small values of the coupling $\lambda$ are straightforward to find starting from the exact expression \eqref{eq:weven}. These series have a finite radius of convergence that is equal to $\widetilde{\lambda}=\pi^2$.
The analysis of the radius of convergence for observables in this quiver gauge theory was first carried out in \cite{Pini:2017ouj} for the free energy and then extended to other observables in \cite{Beccaria:2020hgy,Beccaria:2021hvt}. Expanding the expression \eqref{eq:weven} for small $\lambda$, we obtain:
\begin{equation}\label{eq:20250222a}
\Delta w^{(\alpha)\,\text{even}} \, = \,
-\frac{3\zeta_3}{32 \pi ^4} s_{\alpha}\lambda ^2 
+\frac{\pi ^2 \zeta_3+20 \zeta_5}{256 \pi ^6} s_{\alpha}\lambda ^3 + O\left(  \lambda^4 \right) \,.
 \end{equation}

In Table \ref{table1} we show the comparison between the sum of the first analytical coefficients of the perturbative expansion of \eqref{eq:weven} and the data obtained with the numerical method for the case of the $\mathbb{Z}_3$ quiver and $\alpha=1$. 
\begin{table}[H]
	\centering
	\footnotesize
	\renewcommand{\arraystretch}{1.2}
	$$
	\begin{array}{|c|lllll|}
		\hline
		n \backslash \lambda & 1 & 2 & 3 & 4 & 20 \\
		\hline
1&-0.000867677 & -0.00347071 & -0.00780909 & -0.0138828 & -0.347071 \\
 2&-0.000768326 & -0.0026759 & -0.00512662 & -0.00752438 & 0.447736 \\
 3&-0.000776575 & -0.00280789 & -0.00579479 & -0.00963613 & -0.872106 \\
 \vdots& & &\vdots && \\
 9&-0.000776011 & -0.00279093 & -0.00567341 & -0.0091526 & -3.75858 \\
 10&-0.000776011 & -0.00279093 & -0.0056734 & -0.00915244 & 4.11106 \\
		\hline
		\text{numerics} & -0.000776011&-0.00279093&-0.00567341&-0.00915247&-0.0714413\\
		\hline
	\end{array}
	$$
\caption{Comparison between the numerical method and the perturbative expansion of $\Delta w^{(\alpha)\,\text{even}}$.
The last row gives the value which is calculated with the numerical method. The first row with $n=1$ gives the result using one term in the expansion~\eqref{eq:20250222a}, the second row with $n=2$ gives the result using two terms, and so on. It is clear that for small $\lambda < \widetilde{\lambda}$ ($\lambda = 1,2,3,4$ in the Table) the series~\eqref{eq:20250222a} converges to the numerical result as $n$ increases. While, for large $\lambda > \widetilde{\lambda}$ ($\lambda = 20$ in the Table), the series~\eqref{eq:20250222a} does not converge, but the numerical method is still applicable and provides finite results.}
\label{table1}
\end{table}
We performed a similar test for $\Delta w^{(\alpha)\,\text{odd}}$ and also for many other values of $M$ and $\alpha$. In all cases, the results obtained with our numerical method agreed closely with the small $\lambda$ expansion, providing strong evidence for the correctness of our numerical implementation.
\subsubsection{Large \texorpdfstring{$\lambda$}{} expansion}\label{sec:NumEvidenceM3a1}
\begin{figure}[ht]
\centering
\includegraphics[scale=0.7]{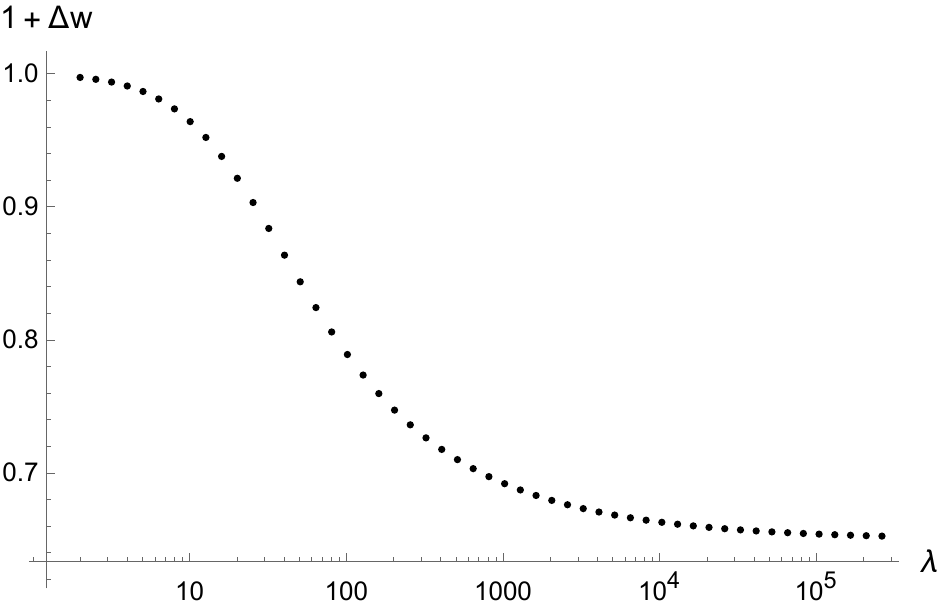}
\caption{$1+\Delta w$ as function of $\lambda$ for $M=3$, $\alpha = 1$. We notice that the function reaches a plateau for values of $\lambda\sim 10^4-10^5$.  A selection of numerical values can be found in Table~\ref{tab:WWM3p1} of Appendix~\ref{app:MoreNumericalEvidence}.}\label{fig1}
\end{figure}

Finally, we provide numerical evidence  supporting the validity of the strong coupling expansion~\eqref{eq:DeltawLargelambda}. As an example, we consider the case with $M=3$, $\alpha=1$. We have also tested for many values of $M$ and $\alpha$, but since the calculations and conclusions are in all cases similar, we prefer not to include them.
For $M=3$, $\alpha = 1$, the expansion~\eqref{eq:DeltawLargelambda} reads
\begin{equation}\label{eq:2pt31}
1 + \Delta w^{(1)}  \underset{\lambda \rightarrow \infty}{\sim} \kappa_0\left(1+  \kappa_1\frac{1}{\sqrt{\lambda}} + \kappa_2 \frac{1}{\lambda}+ \kappa_3 \frac{1}{\lambda^{3/2}} + O\left(\lambda^{-2}\right)\right)
\end{equation}
with
\begin{subequations}    
\begin{align}
\kappa_0 &= \frac{16 \pi ^2}{243} \,,\label{eq:fit31c0}\\
\kappa_1 & = 2\,,\label{eq:fit31c1}\\
\kappa_2 & = 3-\frac{3 \log (3)}{4} \,,\label{eq:fit31c2}\\
\kappa_3 &=-\frac{3}{4} \left(-5+3 \log (3)+3 \log^2(3)\right) \,. \label{eq:fit31c3}
\end{align}
\end{subequations}
We recover these results as follows. We first compute numerical values of $1 + \Delta w^{(1)}$ for $\lambda = 10000, 15000, 20000,\ldots, 300000$, calling this data set $f_{\text{num}}(\lambda)$. We have chosen this range of values because we wanted to start from a sufficiently large value of $\lambda$ for the strong coupling analysis and then we have selected the second endpoint of the interval based on the observation that, as the coupling increases, the function reaches a plateau, as can be seen in Fig.~\ref{fig1}.
Next, we fit the data $f_{\text{num}}(\lambda)$ with the Mathematica function {\tt LinearModelFit[]} using the ansatz 
\begin{align}
\sum_{i=0}^{n}\,f_i\,\lambda^{-\frac{i}{2}}\,.
\label{ansatz}
\end{align}
stopping at $n=6$. We choose this value because it allows us to obtain accurate predictions for the numerical coefficients, while further increasing
$n$ does not lead to any improvement in their precision. 
In this way, we first evaluate the coefficient of the $\lambda^0$ term, finding the numerical estimate of $\kappa_0$ in the expansion~\eqref{eq:2pt31}. The result of this fit is reported in the first row of Table~\ref{table2} and it agrees very well with the closed-form expression \eqref{eq:fit31c0}. Subsequently, to extract the next coefficient $\kappa_1$, we define 
\begin{align}
\tilde f_{\text{num}}(\lambda) = \left( \frac{f_{\text{num}}(\lambda)}{\kappa_0}  - 1\right) \sqrt{\lambda}
\end{align}
and apply {\tt LinearModelFit[]} to the new data set $\tilde f_{\text{num}}(\lambda)$ using the same ansatz \eqref{ansatz}. In this case, the coefficient of the constant term yields an estimate for $\kappa_1$. The result of the fit can be seen in the second row of Table~\ref{table2} and shows excellent agreement with the exact value \eqref{eq:fit31c1}. We proceed similarly to numerically determine $\kappa_2$, as presented in Table~\ref{table2}.
\begin{table}[H]
	\centering
	\renewcommand{\arraystretch}{1.5}
$$
\begin{array}{c@{\hspace{.5cm}}|@{\hspace{1cm}}c@{\hspace{1cm}}c}
& \text{Estimate}  \\
\hline
 \kappa_0 &0.64985049554(4)  \\
 \kappa_1 & 2.00000000(0) \\
\kappa_2 & 2.176040(7) \\
 \end{array}
$$
\caption{The estimates of the first three coefficients in the expansion~\eqref{eq:2pt31}, obtained by fitting the numerical data. 
We have accounted for both model fitting errors and systematic errors, the latter resulting from the discretization process used in the Nyrstr\"om method. 
}
\label{table2}
\end{table}

So far, we have shown that the numerical algorithm accurately reproduces the coefficients in \eqref{eq:2pt31}, which were previously determined analytically. At this point, we can also exploit it to predict higher-order coefficients that we have not computed with analytical methods. In particular, we apply it to estimate $\kappa_3$, leading us to conjecture the closed-form~\eqref{c3}. The latter has been determined in the following way. 

\begin{figure}[ht]
\centering
 \includegraphics[scale=0.9]{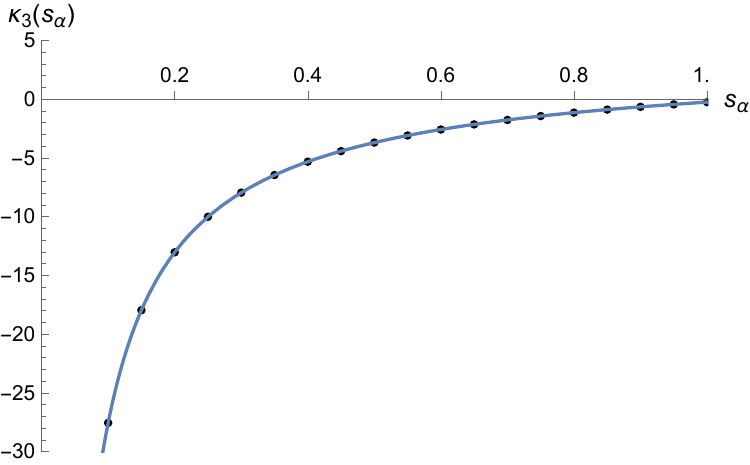}
\caption{Comparison between the numerical data (black dots) for $\kappa_3(s_{\alpha})$ and the quadratic polynomial in $\mI_1(s_{\alpha})$ (blue line) as given by \eqref{Ansatzk3}.  All points are compatible with the fitted function within the numerical errors.}
    \label{fig:331Fitc3}
\end{figure}

The expression for the coefficient of $\lambda^{-3/2}$  in \eqref{eq:2pt31} is strongly constrained by the generating functions reported in Appendix \ref{app:GeneratingFunctions} as well as by the expansion \eqref{w00strong}, which imply that it can only be a polynomial in $ \mathcal{I}_0(s_{\alpha})$ and $ \mathcal{I}_1(s_{\alpha})$ of degree two. Moreover, based on the analytical expressions obtained for the previous coefficients, we further assume that the dependence on $\mathcal{I}_0(s_{\alpha})$ and $s_{\alpha}$ factorizes, with all such dependence being fully encoded in $\kappa_0$. These assumptions lead us to conclude that $\kappa_3$ must be a degree-two polynomial in $ \mathcal{I}_1(s_{\alpha})$, namely
\begin{align}
\sum_{i=0}^{2} g_i \, \mathcal{I}_1(s_{\alpha})^i \,.
\label{Ansatzk3}
\end{align}
The coefficients $g_i$ have been determined by fitting the numerical data for $\kappa_3$ at different values of $s_{\alpha}$ using the function \eqref{Ansatzk3}. The results of this analysis have been reported in Table~\ref{tab:331Fitc3}, while the comparison between the fitted function and the numerical data is shown in Fig.~\ref{fig:331Fitc3}. We regard these results as a strong validation of the expression \eqref{c3}.
\begin{table}[ht]
\renewcommand{\arraystretch}{1.2}
$$
\begin{array}{c|c}
 \text{} & \text{Estimate}   \\
\hline
 g_0 & +3.750(4) \\
 \pi \cdot \,g_1 & -1.500(2)   \\
 \pi^2\cdot g_2 & -0.9999(6)  \\
\end{array}
$$
\caption{Numerical estimates of the coefficients of the quadratic polynomial \eqref{Ansatzk3}. In particular, we observe that, after multiplying by an appropriate power of $\pi$, each coefficient can be identified with a  rational number.}
\label{tab:331Fitc3}
\end{table}
\subsection{The 3-point function \texorpdfstring{$\langle W_{\alpha} W_{\beta} W_{\alpha+\beta}^{\dagger} \rangle$}{}}\label{subsec:3pt}
In this section, we focus on the computation of the large-$\lambda$ expansion of the 3-point function \eqref{ratioWaWbWc}.
As a first step, it is necessary to evaluate the strong coupling expansions of both \eqref{eq:Seven} and \eqref{eq:Sodd}. Here, we simply summarize the results of this  computation. Specifically, for $\mathcal{S}_{\text{even}}^{\alpha}$ we obtained  
\begin{equation}
\mathcal{S}_{\text{even}}^{\alpha} =
c_0^{\text{even}} \sqrt{\lambda} 
+c_1^{\text{even}} + c_2^{\text{even}}\frac{1}{\sqrt{\lambda}}+ 
c_3^{\text{even}}\frac{1}{\lambda}+O\left(\frac{1}{\lambda^{3/2}}\right)
\label{eq:SevenLargelambda}
\end{equation}
with
\begin{subequations}
\begin{align}
c_0^{\text{even}} &= -\frac{1}{2} - \frac{1}{4} \frac{\mI_0(s_{\alpha})}{\sqrt{s_{\alpha}}}\,,
\label{c0even}\\
c_1^{\text{even}} &=-\frac{1}{8}\frac{\mI_0(s_{\alpha})}{\sqrt{s_{\alpha}}}\,,\\
c_2^{\text{even}} &= -\frac{3}{32} \frac{\mI_0(s_{\alpha})}{\sqrt{s_{\alpha}}} - \frac{\pi}{16} \frac{\mI_0(s_{\alpha})}{\sqrt{s_{\alpha}}} \mI_1(s_{\alpha})\,,
\label{c2even}\\
c_3^{\text{even}} &= -\frac{3}{32} \frac{\mI_0(s_{\alpha})}{\sqrt{s_{\alpha}}} - \frac{3 \pi}{32} \frac{\mI_0(s_{\alpha})}{\sqrt{s_{\alpha}}} \mI_1(s_{\alpha})- \frac{ \pi^2}{8} \frac{\mI_0(s_{\alpha})}{\sqrt{s_{\alpha}}} \mI_1(s_{\alpha})^2\label{eq:c3even}\,.
\end{align}
\end{subequations}
Similarly, for $\mathcal{S}_{\text{odd}}^{\alpha}$ we found
\begin{equation}
\mathcal{S}_{\text{odd}}^{\alpha} =c_0^{\text{odd}} \sqrt{\lambda} 
+c_1^{\text{odd}} + c_2^{\text{odd}}\frac{1}{\sqrt{\lambda}}+ 
c_3^{\text{odd}}\frac{1}{\lambda}+O\left(\frac{1}{\lambda^{3/2}}\right)
\label{eq:SoddLargelambda}
\end{equation}
with
\begin{subequations}
\begin{align}
c_0^{\text{odd}} &= -\frac{1}{2} - \frac{1}{4} \frac{\mI_0(s_{\alpha})}{\sqrt{s_{\alpha}}}\,,
\label{c0odd}\\
c_1^{\text{odd}} &=\frac{3}{4}\,,\\
c_2^{\text{odd}} &=  -\frac{3}{16}  +\frac{3 \pi}{16}  \frac{\mI_0(s_{\alpha})}{\sqrt{s_{\alpha}}} \mI_1(s_{\alpha})\,,
\label{c2odd}\\
c_3^{\text{odd}} &= -\frac{3}{16} +\frac{3 \pi}{16} \frac{\mI_0(s_{\alpha})}{\sqrt{s_{\alpha}}} \mI_1(s_{\alpha})+\frac{3 \pi^2}{8} \frac{\mI_0(s_{\alpha})}{\sqrt{s_{\alpha}}} \mI_1(s_{\alpha})^2\label{eq:c3odd}\,.
\end{align}
\end{subequations}
In both the even and the odd cases we determine the first three coefficients using numerical and analytical techniques, whereas we determine the last one only numerically. Substituting the expressions \eqref{eq:SevenLargelambda} and \eqref{eq:SoddLargelambda} into \eqref{ratioWaWbWc}, we then obtain  the large $\lambda$ expansion of $
1+\Delta w^{(\alpha,\beta)}
$, namely
\begin{equation}
1+\Delta w^{(\alpha,\beta)}(M,\lambda) \underset{\lambda \rightarrow \infty}{\sim} c_0
\left[1 + c_1 \dfrac{1}{\sqrt{\lambda}}+c_2\dfrac{1}{\lambda} +c_3\dfrac{1}{\lambda^{3/2}}+ O\left(\frac{1}{\lambda^{2}} \right)\right]
\label{eq:WWWLargelambda}
\end{equation}
with
\begin{subequations}
\begin{align}
c_0& = -\dfrac{1}{8} \prod_{p=1}^3 \dfrac{\mI_0(s_{\alpha_p})}{\sqrt{s_{\alpha_p}}}\,,
\label{c00}\\
c_1 &= 3\,,\\
c_2 &=\frac{1}{4} \left(21 - \pi \sum_{p=1}^3 \mI_1(s_{\alpha_p}) \right)\,,\label{c22}\\
c_3&= \frac{1}{32} \left( 199 - 28 \pi \sum_{p=1}^3 \mI_1(s_{\alpha_p})- 16 \pi^2 \sum_{p=1}^3 \mI_1(s_{\alpha_p})^2\right)\,.
\label{cc3}
\end{align}
\end{subequations}

\subsubsection{Analytical calculation}\label{subsec:WWWanalytical}
As a first step we rewrite the large $\lambda$ expansion of $\mathcal{S}_{\text{even}}^{\alpha}$  \eqref{eq:Seven} and $\mathcal{S}_{\text{odd}}^{\alpha}$  \eqref{eq:Sodd} as
\begin{align}
\mathcal{S}_{\text{even}}^{\alpha} = \sum_{P=0}^{\infty}\frac{\mathcal{S}_{\text{even}}^{(P)}}{\lambda^{P/2}}\, , \qquad  \mathcal{S}_{\text{odd}}^{\alpha} = \sum_{P=0}^{\infty}\frac{\mathcal{S}_{\text{odd}}^{(P)}}{\lambda^{P/2}}\, \ ,        
\end{align}
and we separately consider 
$\mathcal{S}_{\text{even}}^{(P)}$ and $\mathcal{S}_{\text{odd}}^{(P)}$. Then, following the same procedure outlined in Section \ref{subsec:AnalyticSP}, we organize these two quantities as
\begin{align}
& \mathcal{S}_{\text{even}}^{(P)} = \mathcal{S}_{0,\text{even}}^{(P)}+\mathcal{S}_{1,\text{even}}^{(P)}+\mathcal{S}_{2,\text{even}}^{(P)}+ \cdots\, \ , \\[0.5em]
& \mathcal{S}_{\text{odd}}^{(P)} = \mathcal{S}_{0,\text{odd}}^{(P)}+\mathcal{S}_{1,\text{odd}}^{(P)}+\mathcal{S}_{2,\text{odd}}^{(P)}+ \cdots\, \ ,
\end{align}
where $\mathcal{S}_{j,\text{even}}^{(P)}$ encodes the contribution from the coefficients \eqref{wnmCoefficients} $w^{(1)}_{0,s}$ satisfying $s=P-j$, while $\mathcal{S}_{j,\text{odd}}^{(P)}$ the analogous contribution from the coefficients $w^{(2)}_{0,s}$. In particular, after a long computation, we determine
\begin{subequations}
\begin{align}
& \mathcal{S}_{0,\text{even}}^{(P)} = (-2)^{P-2}w_{P,0}^{(1)} \, \ , \\[0.5em]
& \mathcal{S}_{1,\text{even}}^{(P)} = -(-2)^{P-4}(P^2-2P)(1-\delta_{P,0})w_{P-1,0}^{(1)}\, \ , \\[0.5em]
& \mathcal{S}_{2,\text{even}}^{(P)} = -\frac{(-2)^{P-7}}{3} \left(3 P^4-16 P^3+24 P^2-20P+9\right)(1-\delta_{P,0})(1-\delta_{P,1})w_{P-2,0}^{(1)}\, \ ,
\end{align}
\label{SPeven}%
\end{subequations}
while, for the odd sector, we obtain
\begin{subequations}
\begin{align}
& \mathcal{S}_{0,\text{odd}}^{(P)} = (-2)^{P-2}w_{P,0}^{(2)} \, \ , \\[0.5em]
& \mathcal{S}_{1,\text{odd}}^{(P)} = -(-2)^{P-4}(P^2-2P-3)(1-\delta_{P,0})w_{P-1,0}^{(2)}\, \ , \\[0.5em]
& \mathcal{S}_{2,\text{odd}}^{(P)} = -\frac{(-2)^{P-7}}{3}P \left(3 P^3-16 P^2+6 P+16\right)(1-\delta_{P,0})(1-\delta_{P,1})w_{P-2,0}^{(2)}\, \ .
\end{align}
\label{SPodd}%
\end{subequations}
Then, the strong coupling expansions of $ \mathcal{S}_{\text{even}}^{\alpha}$ and $ \mathcal{S}_{\text{odd}}^{\alpha}$ are obtained using \eqref{wStrongCoupling}. In particular, the leading order coefficients, $ c_0^{\text{even}}$ and $c_0^{\text{odd}}$, are explicitly determined in terms of $ S^{(p)}_{0,\text{even}}$ and $ S^{(p)}_{0,\text{odd}}$, respectively and were previously computed in \cite{Pini:2023lyo}. Therefore, let us consider the next-to-leading order and, for simplicity, focus only on the even sector, as the computation for the odd sector can be carried out in a completely analogous way. It turns out that the coefficient $ c_1^{\text{even}}$ is fully determined by the expressions for $\mathcal{S}^{(P)}_{0,\text{even}}$ and $\mathcal{S}^{(P)}_{1,\text{even}}$. After some algebraic manipulations, we obtain
\begin{align}
c_1^{\text{even}} = \frac{1}{4}\sum_{P=0}^{\infty}\frac{\omega_{P,0}^{(1)}}{(-2\pi)^P} - \frac{1}{16}\sum_{P=0}^{\infty}(P^2-2P)\frac{\omega_{P-1,0}^{(1)} }{(-2\pi)^{P}}(1-\delta_{P,0})   
\end{align}
Next, we use the expressions for the generating functions $G^{(1)}(s_{\alpha},\ell,x)$ and $G^{(0)}(s_{\alpha},x)$ collected in Appendix \ref{app:GeneratingFunctions} to rewrite the expression above as  
\begin{align}
c_{1}^{\text{even}} &= \left(\frac{1}{4}G^{(1)}(s_{\alpha},1,x) + \frac{1}{32\pi}\left[(x\partial_x)^2-1\right]G^{(0)}(s_{\alpha},x)\right) \Big|_{x=-2\pi} = -\frac{1}{8} \frac{\mathcal{I}_0(s_{\alpha})}{\sqrt{s_{\alpha}}} \, .
\end{align}  
Using the expressions \eqref{SPodd}, the next-to-leading order term in the odd sector can be derived in a similar manner, yielding  
\begin{align}
c_{1}^{\text{odd}} &= \left(\frac{1}{4}G^{(1)}(s_{\alpha},2,x) + \frac{1}{32\pi}\left[(x\partial_x)^2-4\right]G^{(0)}(s_{\alpha},x)\right) \Big|_{x=-2\pi} = \frac{3}{4} \, .
\end{align}
Finally, the expressions for the next-to-next-to-leading order coefficients, $ c_2^{\text{even}}$ and $c_2^{\text{odd}}$, can be obtained by considering both \eqref{SPeven} and \eqref{SPodd}, and using the results collected in Appendix~\ref{app:GeneratingFunctions}. After a long computation, we find  
\begin{subequations}
\begin{align}
c_{2}^{\text{even}} =& \left(\pi\,G^{(2)}(s_{\alpha},1,x) + \frac{1}{8}((x\partial_x)^2-1)G^{(1)}(s_{\alpha},1,x) \right. \nonumber \\
&\left. + \frac{1}{384\pi}(3(x\partial_x)^4-8(x\partial_x)^3+20(x\partial_x)-15)G^{(0)}(s_{\alpha},x)\right)\Big|_{x=-2\pi}\nonumber \\
=& -\frac{3}{32} \frac{\mathcal{I}_0(s_{\alpha})}{\sqrt{s_{\alpha}}} - \frac{\pi}{16} \frac{\mathcal{I}_0(s_{\alpha})}{\sqrt{s_{\alpha}}} \mathcal{I}_1(s_{\alpha})\, \ , \\[0.5em]
c_{2}^{\text{odd}} =& \left(\pi\,G^{(2)}(s_{\alpha},2,x) + \frac{1}{8}((x\partial_x)^2-4)G^{(1)}(s_{\alpha},2,x) \right. \nonumber \\
&\left. + \frac{1}{384\pi}(3(x\partial_x)^4-8(x\partial_x)^3-18(x\partial_x)^2+56(x\partial_x)-24)G^{(0)}(s_{\alpha},x)\right)\Big|_{x=-2\pi}\, \ \nonumber \\
 =& -\frac{3}{16}  +\frac{3 \pi}{16}  \frac{\mathcal{I}_0(s_{\alpha})}{\sqrt{s_{\alpha}}}\mathcal{I}_1(s_{\alpha}) .
\end{align}
\end{subequations}
We find it important to remark that, although there are no obstructions to analytically compute the coefficients $c_3^{\text{even}}$ and $c_3^{\text{odd}}$, their evaluation turned out to be very involved. In the next section, we show how these results can be recovered and extended numerically.

\subsubsection{Numerical calculation}\label{sec:numericalCurlyS}
For the sake of simplicity, we outline the numerical procedure only for the even case, as the odd case can be treated in a completely analogous manner. As a first step we rewrite
\eqref{eq:Seven} as
\begin{equation}
\mathcal{S}_{\text{even}}^{\alpha}  = - \frac{1}{I_1(\sqrt{\lambda}) }\int_0^{\infty}\!\! dt\ A(t) Z^{(\alpha)}(t)\ ,
\end{equation}  
where we used the relation \eqref{eq3} with $b_\ell = \sqrt{2 \ell}$ and $a_k = \sqrt{2 k}  \,I_{2 k}(\sqrt{\lambda})\,$ and hence $A(t)$ given by \eqref{closedformAandB}. The function $Z^{(\alpha)}(t)$ is the solution of the integral equation
\begin{equation}
Z^{(\alpha)}(t) + s_{\alpha} \int_0^{\infty}\!\!\!\! du\  \KKeven(t,u) Z^{(\alpha)}(u)= -  \frac{s_{\alpha}\,\sqrt{t\,\lambda}}{2\sqrt{2}\pi\,\sinh(t/2)} J_1\left( \frac{t \sqrt{\lambda}}{2 \pi}\right)\, .
\end{equation}
Subsequently, by employing the numerical method described in Section~\ref{sec:NumMethod}, we computed the values of both $\mathcal{S}_{\text{even}}^{\alpha}$ and $\mathcal{S}_{\text{odd}}^{\alpha}$ for various values of the coupling constant $\lambda$, and successfully validated them against their respective weak-coupling expansions.

Following this, we derive the large-$\lambda$ expansions of these quantities using the same approach outlined in Section~\ref{sec:NumEvidenceM3a1} for the 2-point function. We consider many values of $s_{\alpha}$ and $M$, and compute the numerical values\,\footnote{A subset of numerical values can be found in Table~\ref{tab:SevenSodd3p1} of Appendix~\ref{app:MoreNumericalEvidence}.} of $\mathcal{S}_{\text{even}}^{\alpha}$ and $\mathcal{S}_{\text{odd}}^{\alpha}$ for $\lambda = 10000, 15000, 20000, \dots, 200000$. An inspection of the expressions in \eqref{eq:SevenLargelambda} and \eqref{eq:SoddLargelambda} shows that the data must be fitted by the ansatz~\eqref{ansatz}, multiplied by an overall $\sqrt{\lambda}$ factor. Using this procedure, we obtain the numerical values for the first three coefficients of the strong coupling expansions in \eqref{eq:SevenLargelambda} and \eqref{eq:SoddLargelambda}. In all cases, we observe substantial agreement between the analytic predictions and our numerical results, which we interpret as strong evidence supporting the reliability of our numerical method. 

\begin{figure}[ht]
\centering
\includegraphics[scale=0.9]{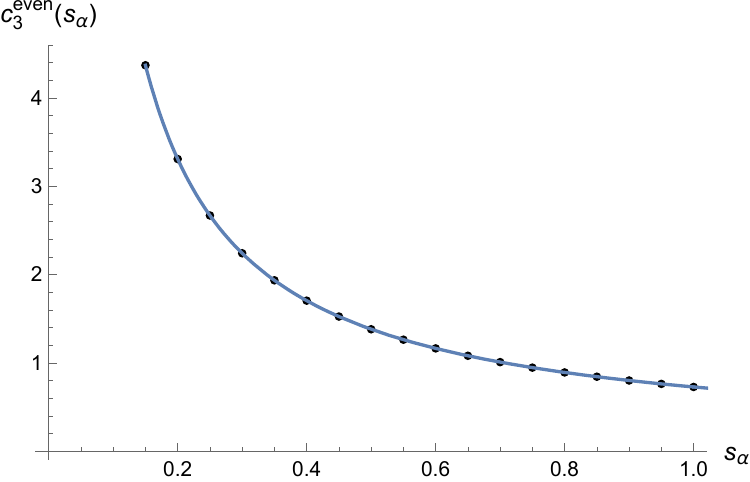}
\caption{Comparison for the even case between the fitted function (blue line), as given by \eqref{c3evenAnsatz}, and the numerical data (black dots) corresponding to different values of the coefficient $c_{3}^{\text{even}}$ for various values of $s_{\alpha}$.  All data points are consistent with the fitted function within numerical errors.}
\label{fig:SevenFitc3}
\end{figure}

\begin{figure}[ht]
\centering
\includegraphics[scale=0.9]{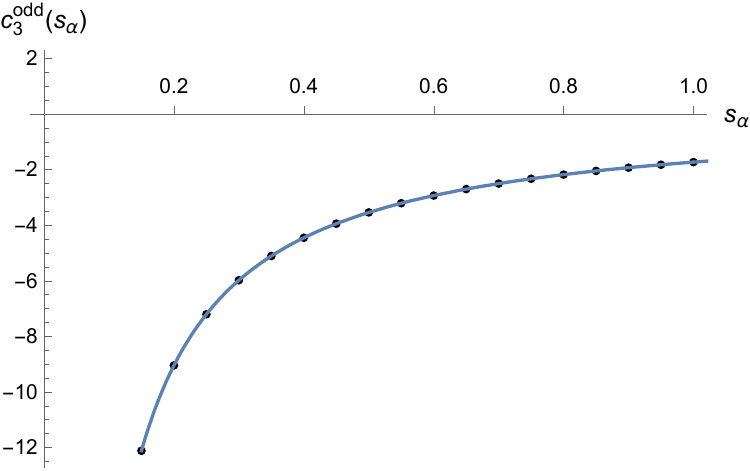}
\caption{Comparison for the odd case between the fitted function (blue line), as given by \eqref{c3oddAnsatz}, and the numerical data (black dots) corresponding to different values of the coefficient $c_{3}^{\text{odd}}$ for varying values of $s_{\alpha}$. All data points are consistent with the fitted function within numerical errors.}
    \label{fig:SoddFitc3}
\end{figure}

Let us now move on to consider the coefficients $c_3^{\text{even}}$ and $c_3^{\text{odd}}$, for which no analytic expressions have been derived. As in the case of the coefficient $\kappa_3$ discussed in Section \ref{sec:NumEvidenceM3a1}, it is important to observe that these two quantities are strongly constrained by the expansion \eqref{w00strong} and the expressions of the generating functions collected in Appendix \ref{app:GeneratingFunctions} which imply that, apart from an explicit dependence on $s_{\alpha}$, they can only be functions of $ \mathcal{I}_0(s_{\alpha})$ and $ \mathcal{I}_1(s_{\alpha})$, with at most a quadratic dependence on $ \mathcal{I}_1(s_{\alpha})$. Furthermore, based on the analytic expressions found for the previous coefficients (see  \eqref{c0even}-\eqref{c2even} and \eqref{c0odd}-\eqref{c2odd}), we find it natural to consider the following ansatz
\begin{subequations}
\begin{align}
c_3^{\, \text{even}} & = \frac{\mathcal{I}_0(s_{\alpha})}{32\,\sqrt{s_{\alpha}}} \sum_{i=0}^{2} c^{\text{even}}_{3,i} \, \left(\pi\,\mathcal{I}_1(s_{\alpha})\right)^i\, \ ,
\label{c3evenAnsatz}
\\[0.5em]
c_3^{\, \text{odd}} & = \frac{c_{3,0}^{\text{odd}}}{16} + \frac{\mathcal{I}_0(s_{\alpha})}{16\,\sqrt{s_{\alpha}}} \sum_{i=1}^{2} c^{\text{odd}}_{3,i} \, \left(\pi\,\mathcal{I}_1(s_{\alpha})\right)^i\, \ .
\label{c3oddAnsatz}
\end{align}
\label{Ansatzc3}
\end{subequations}
The coefficients $c_{3,i}^{\text{even}}$ and $c_{3,i}^{\text{odd}}$ with $i=0,1,2$ have been determined numerically by fitting the values of $c_3^{\text{even}}$ and $c_3^{\text{odd}}$ obtained for different choices of $s_{\alpha}$. The results of this analysis are presented in Table \ref{tab:SevenFitc3}. Remarkably, all the estimates are consistent with integer numbers. This observation allowed us to easily derive the conjectured expressions reported in~\eqref{eq:c3even} and~\eqref{eq:c3odd}.
\begin{table}[ht]
\renewcommand{\arraystretch}{1.2}
$$
\begin{array}{c|c}
\text{} & \text{\shortstack{Estimate}}  \\
\hline
c_{3,0}^{\text{even}}  & -3.000(3) \\[2mm]
c_{3,1}^{\text{even}} & -2.9999(8)  \\[2mm]
c_{3,2}^{\text{even}} & -3.9999(6)  \\[2mm]
c_{3,0}^{\text{odd}}  & -2.999(8)  \\[2mm]
c_{3,1}^{\text{odd}} & 3.000(2)  \\[2mm]
c_{3,2}^{\text{odd}} & 5.9999(4)   \\
\end{array}
$$
\caption{Summary of the numerical values obtained for the coefficients of the ansatz \eqref{Ansatzc3}. Both systematic and  errors from the fitting have been taken into account.}
\label{tab:SevenFitc3}
\end{table}
Moreover, comparisons between the fitted functions and the numerical data are presented in Fig. \ref{fig:SevenFitc3} and Fig. \ref{fig:SoddFitc3} for the even and odd cases, respectively. It is important to observe that, in both cases, all data points lie on the graphs of the fitted functions within numerical precision. We interpret this excellent agreement as strong evidence supporting the validity of our numerical results.

\subsection{The \texorpdfstring{$n$}{}-point function}\label{subsec:npt}
We now discuss the general correlator with $n$ coincident twisted Wilson loops.
Leveraging the properties of Wick's contractions satisfied by the $\mathcal{P}_{\alpha,k}$ operators in the planar limit of the $\mathbb{Z}_M$ quiver gauge theory, it was proven \cite{Pini:2023lyo} that the exact expression of the correlator with $n$ coincident twisted Wilson loops can be derived from the exact results of the correlators with 2 and 3 Wilson loops.

In particular, the case where $n$ is an even number is simply given by the factorization into all possible 2-point correlators and summing over all the possibilities; while, if $n$ is an odd number, the exact expression of the correlator is obtained by the factorization into one 3-point correlator and as many 2-point correlators as needed, and again summing over all the permutations. 

As a first example, let us consider the $\mathbb{Z}_2$ quiver. In this case there is only one twisted sector and, since in each correlator the total ``twist charge'' must be 0 modulo $M$, in this model correlators among an odd number of Wilson loops are vanishing. As shown in \cite{Pini:2023lyo}, in the planar limit the exact expression of the $2n$ Wilson loop correlator reads
\begin{align}
\label{eq:Z2general}
\frac{\big\langle W_1\,W_1\,\dots W_1 \big\rangle}{\left(W_{conn}^{(2)}(\lambda)\right)^n} \simeq \left(1+\Delta w^{(1)}(2,\lambda)\right)^{n} \,.
\end{align}
Hence, using the result obtained in \eqref{eq:DeltawLargelambda}, it is straightforward to find
\begin{align}
\label{eq:Z2generalsc}
\frac{\big\langle W_1\,W_1\,\dots W_1 \big\rangle}{\left(W_{conn}^{(2)}(\lambda)\right)^n} \underset{\lambda \rightarrow \infty}{\sim} &\left( \frac{\pi^2}{16}  \right)^n   \biggl(1\,+\frac{2\,n}{\sqrt{\lambda}} + \frac{n \left( 2n+1-\log 2 \right)}{\lambda} + \notag \\
& + \frac{n \left(16 n^2+24 n \,(1-\log 2)+5-12\log 2\, (1+4\log 2\right)}{12\,\lambda^{3/2}}   + O\left(\lambda^{-2}\right) \biggr) \,.
\end{align}
In the $\mathbb{Z}_M$ quiver theory, let us first consider the correlator with an even number of twisted Wilson loops. Defining the ratio 
\begin{align}
\label{ratioEVEN}
1+\Delta w^{(\alpha_1,\dots,\alpha_{2n-1})} (M,\lambda)  \equiv \frac{\langle\, W_{\alpha_1}\,W_{\alpha_2} \dots W_{\alpha_{2n-1}}W_{\alpha_{2n}}^{\dagger} \,\rangle}{\mathcal{N}^{even}(\alpha_1,\alpha_2,\dots,\alpha_{2n})\,\left(W_{conn}^{(2)}(\lambda)\right)^n} \,,
\end{align}
where $\mathcal{N}^{even}$ is just a normalization factor, in the large $N$ limit we get \cite{Pini:2023lyo}
\begin{align}
\label{evenFinal}
& 1+\Delta w^{(\alpha_1,\dots,\alpha_{2n-1})} (M,\lambda) \,\simeq \,  \nonumber \\
& \frac{1}{\mathcal{N}^{even}(\alpha_1,\alpha_2,\dots,\alpha_{2n})}\left[ \prod_{j=1}^{n}\delta_{\alpha_{2j-1},\alpha_{2j}}\left(1+\Delta w^{(\alpha_{2j-1})}(M,\lambda)\right) + \text{permutations} \right].
\end{align} 
Hence its strong coupling expansion can be immediately obtained by the product of $n$ expansions as \eqref{eq:DeltawLargelambda}. \\
Analogously, considering the ratio of correlators with an odd number of Wilson loops 
\begin{align}
\label{ratioODD}
1+\Delta w^{(\alpha_1,\dots,\alpha_{2n+1})} (M,\lambda)  \equiv \frac{\langle\, W_{\alpha_1}\,W_{\alpha_2} \dots W_{\alpha_{2n}}W_{\alpha_{2n+1}}^{\dagger} \,\rangle}{\mathcal{N}^{odd}(\alpha_1,\alpha_2,\dots,\alpha_{2n+1})\,\sqrt{M}\,W_{conn}^{(3)}(\lambda)\,\left(W_{conn}^{(2)}(\lambda)\right)^{n-1}} \,,
\end{align}
with  $\mathcal{N}^{odd}(\alpha_1,\alpha_2,\dots,\alpha_{2n+1})$ a different normalization factor, the planar exact expression becomes \cite{Pini:2023lyo}
\begin{align}
& 1+\Delta w^{(\alpha_1,\alpha_2,\dots, \alpha_{2n})}(M,\lambda) \,\simeq \, \frac{1}{\mathcal{N}^{odd}(\alpha_1,\alpha_2,\cdots,\alpha_{2n+1})} \nonumber \\
& \left[\delta_{\alpha_1+\alpha_2,\alpha_3}\left(1+\Delta w^{(\alpha_1,\alpha_2)}(M,\lambda)\right)\prod_{j=2}^{n}\delta_{\alpha_{2j},\alpha_{2j+1}}\left(1+\Delta w^{(\alpha_{2j})}(M,\lambda)\right) + \text{permutations} \right] \,.
\label{oddFinal}
\end{align}
Its strong coupling behavior then follows straightforwardly from the product of one expansion as \eqref{eq:WWWLargelambda} and $(n-1)$ expansions as \eqref{eq:DeltawLargelambda}.

\section{Integrated correlator with a Wilson loop}\label{sec:intcorr}
In this section, exploiting the algorithm introduced in Section \ref{sec:NumMethod}, we extend the analysis of \cite{Pini:2024zwi} and compute the strong coupling expansion of the next-to-planar coefficient $\mathcal{W}^{(NL)}$ of \eqref{LargeNW}  up to order $O(\lambda^0)$. We begin by recalling that, for any value of the ’t~Hooft coupling, this quantity is given by  \begin{align} 
\mathcal{W}^{(NL)} = 2\frac{\langle W_0\,\mathcal{M} \rangle_c^{(NL)}}{W^{(L)}} - \mathcal{W}^{(L)}\frac{W^{(NL)}}{W^{(L)}} \ ,
\label{WNLStart}
\end{align}
where $\mathcal{W}^{(L)}$ is given in \eqref{Wleading} 
while $W^{(L)}$ and $W^{(NL)}$ are defined by expressions \eqref{WL} and \eqref{WNL}, respectively. Finally, we recall that the connected correlator $\langle W_0\mathcal{M} \rangle_c^{(NL)}$ can be decomposed as \begin{align} \langle W_0\,\mathcal{M} \rangle_c^{(NL)} = \langle W_0\,\mathcal{M}^{(1)} \rangle_c^{(NL)} + \langle W_0\,\mathcal{M}^{(2)} \rangle_c^{(NL)} \ ,
\end{align} with the two terms on the right-hand side given by expressions (3.7a) and (3.7b) of \cite{Pini:2024zwi}, respectively. 

Henceforth, we focus exclusively on the strong coupling expansion of \eqref{WNLStart}. Building on the analysis presented in \cite{Pini:2024zwi}, and by employing the method introduced in Appendix B of \cite{Pufu:2023vwo}, it is possible to analytically compute the large $\lambda$ expansion of both \eqref{Wleading} and the correlator $\langle W_0 \, \mathcal{M}^{(1)} \rangle_c^{(NL)}$ to any desired order in the ’t~Hooft coupling. For example, for this last quantity, we obtain
\begin{align}
\langle W_0\,\mathcal{M}^{(1)} \rangle_c^{(NL)} \, \underset{\lambda \rightarrow \infty}{\sim} \,  -\frac{\lambda ^2}{768} - \frac{ \lambda^{3/2}}{4608}\left(4\pi^2+45\right) + \frac{\lambda}{2048}  \left(24\pi^2-115-16\log (2)\right) + O(\sqrt{\lambda})\,  .  
\end{align}
On the contrary, the evaluation of $\langle W_0,\mathcal{M}^{(2)} \rangle_c^{(NL)}$ requires more attention, as it consists of the sum of two distinct contributions, namely 
\begin{align}
\frac{I_1(\sqrt{\lambda})}{4\sqrt{2}}\sum_{k,\ell=2}^{\infty} (-)^{ k - k\ell}\, \textsf{M}_{k,\ell} ( \sqrt{k\ \ell} - \textsf{d}_k \textsf{d}_\ell )-(\lambda\partial_{\lambda}\mathcal{F})\sum_{q,p=1}^{\infty}\sqrt{p}\,\textsf{M}_{2p,2q}\sqrt{2q}I_{2q}(\sqrt{\lambda})\,  ,
\label{SecondContribution}
\end{align}
where the functions $\mathsf{d_k}$ were defined in \eqref{dalfa}\,\footnote{Here we are considering the $\mathbb{Z}_2$ quiver where there is only a twisted sector given by $\alpha=1$. For this reason we drop the explicit dependence on $\alpha$ in this section.} and
\begin{align}
\textsf{M}_{k,\ell} = (-1)^{\frac{k+\ell+2k\ell}{2}+1}\sqrt{k\,\ell}\int_0^{\infty}\frac{dt}{t}\frac{(t/2)^2}{\sinh(t/2)^2}\,J_k\left(\frac{t\sqrt{\lambda}}{2\pi}\right)J_{\ell}\left(\frac{t\sqrt{\lambda}}{2\pi}\right) \,  .
\label{Mmatrix}
\end{align}
In particular, while the strong coupling expansion of the term involving the derivative of the free energy can be computed analytically, the first term must be evaluated numerically.  To facilitate its numerical evaluation, we introduce the quantity
\begin{align}\label{defmathcalR}
\mathcal{R} = \sum_{k,\ell=2}^{\infty} (-)^{ k - k\ell}\, \textsf{M}_{k,\ell} ( \sqrt{k\ \ell} - \textsf{d}_k \textsf{d}_\ell )\,  ,    
\end{align}
whose  large $\lambda$ expansion
is detailed in Section \ref{subsec:NumericalEvaluationR}\,\footnote{Previous numerical attempts to evaluate this term, as detailed in Appendix E of \cite{Pini:2024uia}, relied on the construction of a Padé approximant. However, these methods proved less efficient than the numerical approach presented in Section \ref{sec:NumMethod}, as they only provided an estimate for the leading term of the expansion.}. We are forced to use the numerical method because, although the expression for the $\textsf{M}$-matrix in \eqref{Mmatrix} closely resembles that of \eqref{Xmatrix}, the different integration kernel prevents us from expressing \eqref{defmathcalR} in terms of the coefficients \eqref{wnmCoefficients}, thereby precluding the use of the analytical method of Section \ref{sec:WL}\,\footnote{ This does not mean that the strong coupling expansion of this term could not be derived analytically. However, the necessary mathematical techniques have yet to be developed, and although this is undoubtedly an interesting direction, it lies beyond the scope of this work.}. Finally, by including the additional numerical contribution given by \eqref{eq:largelambdaI}, we obtain
\begin{align}
\mathcal{W}^{(NL)}\, \underset{\lambda \rightarrow \infty}{\sim} \, & -\frac{\lambda ^{3/2}}{128} + \frac{\sqrt{\lambda }}{512}\left(8\log(2)-1\right) + \frac{1}{256} \left(2 \zeta_3+32 \log ^2(2) -1 \right) +O(\lambda^{-1/2})\, \ .
\label{WNLStrongExpansion}
\end{align}
This is our final result for the strong coupling expansion of the next-to-leading coefficient $\mathcal{W}^{(NL)}$.

\subsection{Numerical evaluation of \texorpdfstring{$\mathcal{R}(\lambda)$}{}}
\label{subsec:NumericalEvaluationR}
In this section we numerically evaluate the first orders of the large $\lambda$ expansion of the function $\mathcal{R}(\lambda)$ defined in \eqref{defmathcalR}, with the result being
\begin{equation}
\label{eq:largelambdaI}
\mathcal{R}(\lambda) = -\frac{\sqrt{\lambda}}{24} + \frac{5}{8} \frac{1}{\sqrt{\lambda}}+ \frac{\log(2)}{2\,\lambda}+ O\left(\frac{1}{\lambda^{3/2}}\right)\, .
\end{equation}
Let us describe the details of the  computation. As a first step we split the sum \eqref{defmathcalR} in even and odd contributions: $\mathcal{R} = \mathcal{R}_1 + \mathcal{R}_2+\mathcal{R}_3+\mathcal{R}_4$, where
\begin{align}
\mathcal{R}_1(\lambda) &= \
\sum_{k,\ell=1}^{\infty} \sfM_{2k, 2\ell} \left( \sqrt{(2 k)\ (2 \ell)} - \sfd_{2k} \sfd_{2\ell} \right)\,,\label{eq:defcalI1}\\
\mathcal{R}_2(\lambda) &= \sum_{k,\ell=1}^{\infty} \sfM_{2 k, 2 \ell +1} \left( \sqrt{2 k\ (2 \ell+1)} - \sfd_{2k} \sfd_{2\ell+1} \right)\,,\\
\mathcal{R}_3(\lambda) &= -\sum_{k,\ell=1}^{\infty} \sfM_{2 k +1, 2 \ell} \left( \sqrt{(2 k+1)\  2\ell} - \sfd_{2k+1} \sfd_{2\ell} \right)\,,\\
\mathcal{R}_4(\lambda) &= \sum_{k,\ell=1}^{\infty} \sfM_{2 k +1, 2 \ell+1} \left( \sqrt{(2 k+1)\  (2\ell+1)} - \sfd_{2k+1} \sfd_{2\ell+1} \right)\,.
\end{align}
We observe that, since \eqref{Mmatrix} is a symmetric matrix, it follows that 
$\mathcal{R}_2(\lambda)+\mathcal{R}_3(\lambda)=0$, so we only need to consider $\mathcal{R}_1(\lambda)$ and $\mathcal{R}_4(\lambda)$. For simplicity, we outline only the numerical evaluation of the even contribution corresponding to $\mathcal{R}_1(\lambda)$, as the odd contribution can be derived in a completely similar way. We begin by observing that the coefficient \eqref{dalfa} with $s_{\alpha}=1$ can be obtained by multiplying both sides of equation \eqref{eq1} by $\sqrt{2\ell}$ and then performing the sum over 
$\ell$, namely
\begin{align}
\textsf{d}_{2k} = \sqrt{2k} -  \int_0^{\infty} dt\, U_k^{\text{even}}(t) \,Z(t)\, ,
\label{eq:dk}
\end{align}
where $U_{k}^{\text{even}}$ was defined in \eqref{eq:Vk} and $Z(t)$ is the solution of the integral equation
\begin{align}
Z(t) + \int_0^{\infty}du\, \KKeven(t,u)Z(u) = - \frac{\sqrt{t\,\lambda}}{2\sqrt{2}\pi\,\sinh(t/2)}J_1\left(\frac{t\sqrt{\lambda}}{2\pi}\right)\, .     
\end{align}
Then, by employing the expression \eqref{eq:dk}, it follows that \eqref{eq:defcalI1} can be rewritten as
\begin{align}
\mathcal{R}_1(\lambda) = & \, 2\sum_{k,\ell=1}^{\infty}\sqrt{2k}\,\textsf{M}_{2k,2\ell}\int_0^{\infty}dt\, U_{\ell}^{\text{even}}(t)Z(t) \nonumber \\
& - \sum_{k,\ell=1}^{\infty}\textsf{M}_{2k,2\ell}\int_0^{\infty}dt\,\int_0^{\infty} dt'\, U_{k}^{\text{even}}(t)\,U_{\ell}^{\text{even}}(t')Z(t)\,Z(t') \, .
\label{R1Finale}
\end{align}
Finally, we observe that by using the explicit form of the matrix $\textsf{M}$ given in \eqref{Mmatrix}, together with the properties of the Bessel functions, the sums over $k$ and $\ell$ in \eqref{R1Finale} can be performed analytically, leading to an expression that can be efficiently evaluated numerically.

Using the method outlined in Section \ref{sec:NumMethod} we evaluate $\mathcal{R}_1(\lambda)$ and $\mathcal{R}_4(\lambda)$ numerically for different values of the coupling.  As a first step, we validate the numerical data by comparing them with the corresponding weak coupling expansions, finding perfect agreement. We then proceed to analyze their behavior in the strong coupling limit. We find it useful to consider directly the sum $\mathcal{R}(\lambda)$, whose large $\lambda$ expansion is of the form
\begin{align}
\mathcal{R}(\lambda) \underset{\lambda \rightarrow \infty}{\sim} \sqrt{\lambda}\, \widetilde{\mathcal{R}}_{1} + \widetilde{\mathcal{R}}_2 + \frac{\widetilde{\mathcal{R}}_3}{\sqrt{\lambda}} + \frac{\widetilde{\mathcal{R}}_4}{\lambda} + O\left(\frac{1}{\lambda^{3/2}}\right)\, .
\label{LargeR}
\end{align}
To determine the coefficients $\widetilde{\mathcal{R}}_i$, we follow the same procedure outlined in Section \ref{sec:NumEvidenceM3a1}. Specifically, we first evaluate $\mathcal{R}(\lambda)$ for  $\lambda = 10000, 15000, 20000,\ldots, 500000$. We then fit the numerical data\footnote{A subset of the numerical data can be found in Table~\ref{tab:calI1and4} in Appendix~\ref{app:MoreNumericalEvidence}.} using the ansatz \eqref{ansatz} multiplied by $\sqrt{\lambda}$. The results of this analysis are reported in Table \ref{tab:IntegratedCorrFits}, and based on these findings, we conjecture the large $\lambda$ expansion \eqref{eq:largelambdaI}. Finally, we observe that, in principle, the subleading coefficients $\widetilde{\mathcal{R}}_{i \geq 5}$ could also be determined. However, the numerical accuracy is insufficient to support the formulation of any closed-form conjecture for these quantities, and therefore we choose not to report them here.

\begin{table}[H]
	\centering
	\renewcommand{\arraystretch}{1.5}
$$
\begin{array}{c@{\hspace{1cm}}|@{\hspace{1cm}}c@{\hspace{1cm}}c}
\text{} & \text{Estimate}  \\
\hline
\widetilde{\mathcal{R}}_1 & 	-0.041666666666(6)\\
\widetilde{\mathcal{R}}_2 & -0.00000000000(5)\\
\widetilde{\mathcal{R}}_3 &  0.625000000(2)\\
\widetilde{\mathcal{R}}_4 &  0.346573(8)\\
 \end{array}
$$
\caption{Summary of the numerical values obtained for the coefficients of the ansatz \eqref{LargeR}. Both systematic and errors from the fitting have been taken into account.}
\label{tab:IntegratedCorrFits}
\end{table}
\textcolor{blue}{
}
\section{Conclusions}
\label{sec:conclusions}
In this paper, within the context of the large $N$ limit of the $4d$ $\mathcal{N}=2$ quiver gauge theory at the orbifold fixed point, we computed the first subleading terms in the strong coupling expansions of different correlators involving the insertion of Wilson line operators. Specifically, we determined the large $\lambda$ expansion 
for the planar terms of 2-point and 3-point correlators among coincident twisted Wilson loops, which are given by expressions \eqref{eq:DeltawLargelambda} and \eqref{eq:WWWLargelambda} respectively. Then, as discussed in Section \ref{subsec:npt}, starting from these results, the strong coupling expansion of the generic $n$-point correlator can be immediately derived. Furthermore, we obtained the strong coupling expansion \eqref{WNLStrongExpansion} for the next-to-planar term of the integrated correlator of two moment map operators in the presence of a Wilson line \eqref{intcorrW}. 

We observe that, in analogy to what has been done in \cite{Beccaria:2022ypy,Korchemsky:2025eyc}, we could rewrite the expansions \eqref{eq:DeltawLargelambda} and \eqref{eq:WWWLargelambda} by rescaling the 't Hooft coupling as follows
\begin{align}
 \lambda \mapsto \lambda' \equiv \lambda -4\pi\mathcal{I}_1(s_{\alpha})\sqrt{\lambda} +4\pi^2\mathcal{I}_1(s_{\alpha})^2\,  ,
 \label{lambdaprime}
\end{align}
where $\mathcal{I}_1(s_{\alpha})$ was defined in \eqref{I1}. From a physical point of view, the redefinition \eqref{lambdaprime} shows that the effective coupling $\lambda'$ depends non-trivially on $s_{\alpha}$ and therefore on the twisted sector. Since the $\mathcal{N}=2$ quiver gauge theory has a known gravity dual, we find it interesting to investigate in the future what this redefinition of the coupling implies from the holographic perspective. 

Furthermore, it is important to emphasize that the observables we have considered in this work are just examples of quantities which cannot be written as Fredholm determinants of Bessel operators. In fact, both the analytical and numerical methods we have outlined can have a wide range of applications. Indeed, the expressions for the generating functions collected in Appendix \ref{app:GeneratingFunctions}, and used for analytical calculations in Section \ref{sec:WL}, do not depend on the matrix model representation of a specific operator and can potentially be applied whenever the large $\lambda$ expansion of a given observable can be expressed in terms of the coefficients \eqref{wnmCoefficients}. Moreover, the numerical method outlined in Section \ref{sec:NumMethod} can be also employed in contexts where analytical methods are not currently available. For instance, in the context of the 
$\mathcal{N}=2$ quiver gauge theory, it would be valuable to explore how these techniques can be applied to derive the strong coupling expansion of the next-to-planar contributions to extremal two- and three-point correlators between chiral and antichiral scalar operators. These techniques can also be applied to a broader class of gauge theories, among which natural candidates are the \textbf{E} theory, for which exact expressions for many observables have been obtained over the last few years (see for instance \cite{Beccaria:2020hgy,Beccaria:2021hvt,Billo:2022xas}), as well as the \textbf{D} theory recently considered in \cite{Beccaria:2021ism,Billo:2024ftq}. In this context, a particularly interesting quantity is the integrated correlator of two moment map operators and two Coulomb branch operators \cite{Binder:2019jwn}. Importantly in $\mathcal{N}=2$ theories \cite{Billo:2023kak,Pini:2024uia} only the leading term of its strong coupling expansion is currently known. It is therefore worth investigating whether the subleading contributions can be analyzed using the techniques presented in this work.

\section*{Acknowledgments}
We would like to thank M. Billò, N. Bobev, L. De Lillo, M. Frau and A. Lerda for many important discussions. PJDS is supported in part by FWO projects G003523N, G094523N, and G0E2723N, as well as by the Odysseus grant G0F9516N from the FWO. The work of AP is supported  by the Deutsche Forschungsgemeinschaft (DFG, German Research Foundation) via the Research Grant ``AdS/CFT beyond the classical supergravity paradigm: Strongly coupled gauge theories and black holes” (project number 511311749). The work of PV is partially supported by the MUR PRIN contract 2020KR4KN2 “String Theory as a bridge between Gauge Theories and Quantum Gravity” and by the INFN project ST\&FI “String Theory \& Fundamental Interactions”.
\appendix

\section{Derivation of the generating functions}
\label{app:GeneratingFunctions}
In this appendix we provide a derivation of the generating functions used in the analytic strong coupling computations performed in Section \ref{sec:WL}. To this regard we first recall that the generating functions for the coefficients \eqref{wnmCoefficients} are defined as \cite{Beccaria:2023kbl,Pini:2023lyo}
\begin{subequations}
\begin{align}
G(s_{\alpha},\ell,x) &= \sum_{n=0}^{\infty}\frac{w_{0,n}^{(\ell)}}{(gx)^n} = g\,G^{(0)}(s_{\alpha},x) + G^{(1)}(s_{\alpha,}\ell,x)+g^{-1}\,G^{(2)}(s_{\alpha},\ell,x)+ \cdots \,  , \label{DefG}\\
G(s_{\alpha},\ell,x,y) &= \sum_{n=0}^{\infty}\sum_{m=0}^{\infty}\frac{w_{n,m}^{(\ell)}}{(gx)^n(gy)^m} = g\,G^{(0)}(s_{\alpha},x,y) + G^{(1)}(s_{\alpha,}\ell,x,y) + \cdots \, .\label{DefGG}
\end{align}
\label{GeneratingFunctions}%
\end{subequations}
It is important to treat separately the generating function for the coefficients $w_{0,n}^{(\ell)}$ and that for $w_{n,m}^{(\ell)}$. Since, as demonstrated in \cite{Belitsky:2020qir,Beccaria:2023kbl,Pini:2023lyo}, it can be shown that all the coefficients $w_{n,m}^{(\ell)}$ can be recovered starting from the coefficients $w_{0,n}^{(\ell)}$.  
For this reason, we will focus solely on these in the following. Their expansion at strong coupling can be easily obtained by employing \eqref{wStrongCoupling} and it reads
\begin{align}
w_{0,n}^{(\ell)}  = \sum_{i\geq 0}\omega_{0,n}^{(\ell,i)}\,g^{n+1-i}\,  ,
\label{wExpansion}
\end{align}
with $g$ defined in \eqref{gCoupling}. The case $n = 0$ has been thoroughly analyzed in \cite{Belitsky:2020qir,Belitsky:2020qrm,Beccaria:2022ypy,Pini:2023lyo}, where it was shown how the corresponding strong coupling expansion can be determined to any desired order in $ g$. In particular, the first terms read
\begin{align}
w_{0,0}^{(\ell)} & = 4g\mathcal{I}_{0}(s_{\alpha}) + (2\ell-1) -\frac{(2\ell-1)(2\ell-3)\mathcal{I}_1(s_{\alpha})}{8g}- \frac{(2\ell-1)(2\ell-3)\mathcal{I}_1(s_{\alpha})^2}{16g^2} \nonumber \\
& -\frac{(2\ell-1)(2\ell-3)(16\mathcal{I}_1(s_{\alpha})^3-5\mathcal{I}_2(s_{\alpha})-8\mathcal{I}_2(s_{\alpha})\ell+4\mathcal{I}_2(s_{\alpha})\ell^2)}{512g^3} + O\left(\frac{1}{g^4}\right)\, ,
\label{w00strong}
\end{align}
where $\mathcal{I}_{n}(s_{\alpha})$ was defined in \eqref{eq:defCurlyI2308.03848}. Then, in \cite{Pini:2023lyo}, it was shown that the leading coefficients of the strong coupling expansions of $w_{0,n}^{(\ell)}$ are completely determined by $ \omega_{0,0}^{(\ell,0)}$, via the following recursive relations
\begin{subequations}
\begin{align}
\omega^{(\ell,0)}_{0,2n+1}(s_\alpha) &= - \frac{1}{2(n+1)} \sum_{j=0}^{n} (-1)^{j} \mathcal{I}_{-j}(s_\alpha) \omega^{(\ell,0)}_{0,2n-2j}(s_\alpha)\,, \\
\omega^{(\ell,0)}_{0,2n}(s_\alpha) &= \frac{1}{2n + 1}\left( \sum_{j=1}^{n} (-1)^{n-j-1} \mathcal{I}_{-n+j}(s_\alpha) \omega^{(\ell,0)}_{0,2j-1}(s_\alpha) + 4(-1)^n \mathcal{I}_{-n}(s_\alpha) \right)\,  ,
\end{align}
\label{RecursionEquations}
\end{subequations}
with $n\geq 1$. Starting from these relations the explicit expression for the generating function $G^{(0)}(s_{\alpha},\ell,x)$ can be subsequently worked out as explicitly shown in Appendix B.4 of \cite{Pini:2023lyo}.

Crucially, by using the method developed in \cite{Belitsky:2020qrm,Belitsky:2020qir}, the subleading coefficients $\omega^{(\ell,i)}_{0,n}$  for $i \geq 1$ of the expansion \eqref{wExpansion} can also be determined. Specifically, it turns out that the coefficients $\omega_{0,n}^{(\ell,i)}$ depend solely on $\omega_{0,0}^{(\ell,j)}$ with $j \leq i$ and can be constructed from this finite set of contributions using a system of recurrence relations analogous to the expressions \eqref{RecursionEquations}. For example, for $i = 1$ and the first values of $n$, one can show that 
\begin{subequations}
\begin{align}
\omega_{0,0}^{(\ell,1)} &= 2\ell-1\, , \\[0.5em]
\omega_{0,1}^{(\ell,1)} &= -(\omega_{0,0}^{(\ell,1)}+1)\mathcal{I}_{0}(s_{\alpha}) \, , \\[0.5em]
\omega_{0,2}^{(\ell,1)} &=- \left[\frac{\omega_{0,0}^{(\ell,0)}}{8} -\frac{1}{2}(\omega_{0,0}^{(\ell,1)}+1)\mathcal{I}_0(s_{\alpha})\right]\mathcal{I}_0(s_{\alpha})\, ,
\end{align}
\end{subequations}
and the other coefficients $\omega_{0,n}^{(\ell,1)}$ with $n\geq 3$ can be easily computed as well. Notably, by analyzing these expressions and further using \eqref{RecursionEquations}, we obtain the following closed-form for the coefficients $\omega_{0,n}^{(\ell,1)}$
\begin{align}
\omega_{0,n}^{(\ell,1)} = 
\begin{cases}
2\ell - 1, & n = 0\,, \\
\left[-\frac{\ell}{2} + \left(\frac{n-1}{2}\right)^2\right] \omega^{(\ell,0)}_{0,n-1}, & n \geq 1\,.
\end{cases}
\label{CloseOmegai1}
\end{align}
Therefore, using \eqref{CloseOmegai1} and the definition \eqref{DefG}, the generating function $G^{(1)}(s_{\alpha},\ell,x)$ takes the following form
\begin{align}
 G^{(1)}(s_{\alpha},\ell,x) & = 
\sum_{n=0}^{\infty}\frac{\omega_{0,n}^{(\ell,1)}}{x^n} = \omega_{0,0}^{(\ell,1)} + \frac{1}{4x}\sum_{n=1}^{\infty}\left[(n-1)^2-2\ell\right]\frac{\omega_{0,n-1}^{(\ell,0)}}{x^{n-1}} \nonumber \\
 & = 2\ell-1 + \frac{1}{4x}\left[-2\ell+(x\partial_x)^2\right]G^{(0)}(s_{\alpha},x)\, .
\label{G1}
\end{align}

Following the same procedure, a closed-form expression for the coefficients $\omega_{0,n}^{(\ell,2)}$ can also be derived. After a long calculation, we obtain
\begin{align}
\omega_{0,n}^{(\ell,2)} = 
\begin{cases}
-\frac{(2\ell-1)(2\ell-3)\mathcal{I}_1(s_{\alpha})}{8}, & n = 0\,, \\[0.5em]
\left(\frac{3}{8}-\ell+\frac{\ell^2}{2}\right)\mathcal{I}_0(s_{\alpha})\mathcal{I}_1(s_{\alpha}) -\frac{\ell^2}{2}-\frac{\ell}{2}+\frac{3}{8}, & n =1\,, \\[0.5em]
\left(\frac{3}{4}-2\ell+\ell^2\right)\frac{\mathcal{I}_1(s_{\alpha})\,\omega_{0,n-1}^{(\ell,0)}}{8} + \left(n\ell^2-n(n-2)\ell+\frac{n(n-2)(20-14n+3n^2)}{12}\right)\frac{\omega_{0,n-2}^{(\ell,0)}}{8}, & n \geq 2\,.
\end{cases}
\label{CloseOmegai2}
\end{align}
Then using \eqref{CloseOmegai2} and the definition \eqref{DefG} we obtain the corresponding generating function
\begin{align}
& G^{(2)}(s_{\alpha},\ell,x) = \sum_{n=0}^{\infty}\frac{\omega_{0,n}^{(\ell,2)}}{x^n} = -\frac{(2\ell-1)(2\ell-3)\mathcal{I}_1(s_{\alpha})}{8} + \frac{1}{x}\left(\frac{3}{8}-\frac{\ell}{2}-\frac{\ell^2}{2}\right) + \nonumber \\
&\left(\frac{3}{4}-2\ell+\ell^2\right)\frac{\mathcal{I}_1(s_{\alpha})}{8x}G^{(0)}(s_{\alpha},x) + \frac{1}{8x^2}\left[\ell^2(2-x\partial_{x})-\ell((x\partial_x)^2-2x\partial_x) + \nonumber \right.\\
&\left. \frac{1}{4}(x\partial_x)^4 - \frac{1}{3}(x\partial_x)^3 - \frac{2}{3}(x\partial_x)\right]G^{(0)}(s_{\alpha},x)\,  .
\label{G2}
\end{align}
It is important to observe that both \eqref{G1} and \eqref{G2} are given by a differential operator acting solely on $G^{(0)}(s_{\alpha},x)$. In principle, following the same procedure, the expressions for $G^{(i)}(s_{\alpha},\ell,x)$ with $i \geq 3$ can also be derived.

Let us now consider the generating functions \eqref{DefGG} associated with the coefficients $w^{(\ell)}_{n,m}$. In this regard, it is important to recall that, as shown in \cite{Pini:2023lyo}, they are related to \eqref{DefG} by the following equation
\begin{align}
(gx+gy+1)G(s_{\alpha},\ell,x,y)+\frac{1}{2}(x\partial_x+y\partial_y-g\partial_g)G(s_{\alpha},\ell,x,y) & =  -\frac{1}{4}G(s_{\alpha},\ell,x)G(s_{\alpha},\ell,y) + \nonumber\\
& gx\,G(s_{\alpha},\ell,y)+gy\,G(s_{\alpha},\ell,x)\, \ .
\label{EquationGandGG}
\end{align}
Then, by inserting the definitions \eqref{GeneratingFunctions} into \eqref{EquationGandGG}, we can systematically derive the expressions for the generating functions $G^{(i)}(s_{\alpha},\ell,x,y)$, provided that the expressions for $G^{(i)}(s_{\alpha},\ell,x)$ are known. We recall that, the case $i=0$, yields to \cite{Pini:2023lyo}
\begin{align}
(x+y)G^{(0)}(s_{\alpha},x,y) = -\frac{1}{4}G^{(0)}(s_{\alpha},x)G^{(0)}(s_{\alpha},y)+x\,G^{(0)}(s_{\alpha},y)+y\,G^{(0)}(s_{\alpha},x)\,  ,
\label{GG0}
\end{align}
which allows us to readily obtain the expression for $G^{(0)}(s_{\alpha},x,y)$. 
While, after some algebra, for $i=1$ we obtain  
\begin{align}
4(x+y)G^{(1)}(s_{\alpha},x,y) \, =  & \, (4y-G^{(0)}(s_{\alpha},y))G^{(1)}(s_{\alpha},\ell,x) + (4x-G^{(0)}(s_{\alpha},x))G^{(1)}(s_{\alpha},\ell,y) \nonumber  \\
& -2(G^{(0)}(s_{\alpha},x,y)+x\partial_xG^{(0)}(s_{\alpha},x,y)+y\partial_yG^{(0)}(s_{\alpha},x,y))\, .
\label{GG1equation}
\end{align}
Then, using both \eqref{G1} and \eqref{GG0}, as a final step, we can express $G^{(1)}(s_{\alpha},\ell,x,y)$ solely as a function of $G^{(0)}(s_{\alpha},x)$, namely  
\begin{align}
& G^{(1)}(s_{\alpha},\ell,x,y) = -\frac{x\,G^{(0)}(s_{\alpha},y)\partial_{x}^2G^{(0)}(s_{\alpha},x)}{16 (x+y)}+\frac{x
y \,\partial_{x}^2G^{(0)}(s_{\alpha},x)}{4 (x+y)}+\frac{x y\partial_{y}^2G^{(0)}(s_{\alpha},y)}{4(x+y)} \nonumber \\
& -\frac{y G^{(0)}(s_{\alpha},x)\partial_{y}^2G^{(0)}(s_{\alpha},y)}{16 (x+y)}+\frac{xG^{(0)}(s_{\alpha},y)\partial_{x}G^{(0)}(s_{\alpha},x)}{8 (x+y)^2}-\frac{x y \partial_{x}G^{(0)}(s_{\alpha},x)}{2
(x+y)^2}+\frac{x \partial_{y}G^{(0)}(s_{\alpha},y)}{4 (x+y)} \nonumber \\
& -\frac{x y \partial_{y}G^{(0)}(s_{\alpha},y)}{2 (x+y)^2}-\frac{G^{(0)}(s_{\alpha},y)\partial_{x}G^{(0)}(s_{\alpha},x)}{16 (x+y)}+\frac{y\partial_{x} G^{(0)}(s_{\alpha},x)}{4(x+y)}-\frac{G^{(0)}(s_{\alpha},x) \partial_{y}G^{(0)}(s_{\alpha},y)}{16
   (x+y)}\nonumber \\
& +\frac{yG^{(0)}(s_{\alpha},x) \partial_{y}G^{(0)}(s_{\alpha},y)}{8
   (x+y)^2}-\frac{\ell x G^{(0)}(s_{\alpha}y)}{2 y (x+y)}-\frac{\ell
   G^{(0)}(s_{\alpha},x)}{2 (x+y)}+\frac{\ell G^{(0)}(s_{\alpha},x)
   G^{(0)}(s_{\alpha},y)}{8 y (x+y)} \nonumber\\
   &-\frac{\ell\,G^{(0)}(s_{\alpha},y)}{2
   (x+y)}-\frac{\ell y G^{(0)}(s_{\alpha},x)}{2 x (x+y)}+\frac{\ell
   G^{(0)}(s_{\alpha},x)G^{(0)}(s_{\alpha},y)}{8 x (x+y)}+\frac{x^2
   G^{(0)}(s_{\alpha},y)}{2 (x+y)^3}+\frac{y^2 G^{(0)}(s_{\alpha},x)}{2
   (x+y)^3} \nonumber \\
& +\frac{x y G^{(0)}(s_{\alpha},x)}{2 (x+y)^3}-\frac{x
   G^{(0)}(s_{\alpha},x) G^{(0)}(s_{\alpha},y)}{8 (x+y)^3}-\frac{x
   G^{(0)}(s_{\alpha},y)}{(x+y)^2}+\frac{x y G^{(0)}(s_{\alpha},y)}{2
   (x+y)^3}+\frac{G^{(0)}(s_{\alpha},x)}{4 (x+y)} \nonumber \\
& -\frac{y
   G^{(0)}(s_{\alpha},x)}{(x+y)^2}+\frac{G^{(0)}(s_{\alpha},x)
   G^{(0)}(s_{\alpha},y)}{8 (x+y)^2}-\frac{y G^{0}(s_{\alpha},x)
   G^{(0)}(s_{\alpha},y)}{8 (x+y)^3}+\frac{G^{(0)}(s_{\alpha},y)}{4
   (x+y)}+\frac{2 \ell x}{x+y} \nonumber\\
& +\frac{2 
   \ell\,y}{x+y}-\frac{x}{x+y}-\frac{y}{x+y}\,  .
\label{GG1}   
\end{align}
Let us finally analyze the $i=2$ case. This time the relation \eqref{EquationGandGG} gives 
\begin{align}
-4(x+y)G^{(2)}(s_{\alpha},\ell,x,y) \,= & \, G^{(1)}(s_{\alpha},\ell,x)G^{(1)}(s_{\alpha},\ell,y) + (G^{(0)}(s_{\alpha},x)-4x)G^{(2)}(s_{\alpha},\ell,y)  \nonumber \\
& + (G^{(0)}(s_{\alpha},y)-4y)G^{(2)}(s_{\alpha},\ell,x) + 4G^{(1)}(s_{\alpha},\ell,x,y) \nonumber \\
& + 2x\partial_xG^{(1)}(s_{\alpha},\ell,x,y) + 2y\partial_yG^{(1)}(s_{\alpha},\ell,x,y)\,  .
\end{align}
Then, using \eqref{G1}, \eqref{G2}, as well as \eqref{GG0} and \eqref{GG1},  we obtain an expression for $G^{(2)}(s_{\alpha},\ell,x,y)$ solely in terms of $G^{(0)}(s_{\alpha},x)$ and its derivatives\,\footnote{Since it is very long, we prefer not to report such an expression here; however, we have used it in performing the analytic computations in Section \ref{sec:WL}.}.

\section{Numerical data}\label{app:MoreNumericalEvidence}
In this appendix we collect a subset of the numerical data that have been used in performing the evaluations of the correlators with 2 and 3 coincident Wilson loops in Section \ref{sec:WL} and the computation of the integrated correlator in Section \ref{sec:intcorr}. We think that these data might be useful for a reader interested in reproducing our numerical calculations.

\begin{table}[H]
	\centering
	\footnotesize
	\renewcommand{\arraystretch}{1.2}
	$$
	\begin{array}{|c|c|}
		\hline
		\lambda & \Delta w \\
\hline
1&-0.0007765467702\\
10&-0.03565805023\\
100&-0.2099764972\\
1000&-0.3076811957\\
10000&-0.3370121618\\
100000&-0.346025379\\
200000&-0.347236225\\
300000&-0.3477718781\\
\hline
\end{array}
$$
\caption{$\Delta w$ with $M=3$, $\alpha = 1$: values obtained with the numerical method from Section~\ref{sec:230803848NumMethod}. We used the following choice of cut-off $L$ and number of discretisation points $m$: for $\lambda = 1$: $L=33$, $m=50$; for $\lambda = 10$: 
    $L =30$, $m=50$; for $\lambda = 100$: $L =20$, $m=100$; for $\lambda = 10000$: $L =27$, $m=550$; for $\lambda = 100000$: $L =27$, $m=1250$; for $\lambda = 300000$: $L =29$, $m=2500$. By changing the values of $L$ and $m$, we estimate that this leads to around 14 digits precision.}\label{tab:WWM3p1}
\end{table}
\begin{table}[H]
	\centering
	\footnotesize
	\renewcommand{\arraystretch}{1.2}
	$$
	\begin{array}{|c|c|c|}
		\hline
		\lambda & \mathcal{S}_{\text{even}}^{\alpha}& \mathcal{S}_{\text{odd}}^{\alpha} \\
\hline
1& -3.815846875\times 10^{-4}&-1.607810212\times 10^{-6}\\
10&-4.976539059\times 10^{-2}&-3.996343634\times 10^{-3}\\
100&-7.170174666\times 10^{-1}&-3.319280512\times 10^{-1}\\
1000&-2.852619572&-2.339694719\\
10000&-9.488542748&-8.950447264\\
100000&-3.045051016\times 10^1&-2.990524799\times 10^1\\
200000&-4.314772036\times 10^1&-4.260151265\times 10^1\\
\hline
	\end{array}
	$$
	\caption{Values obtained with the numerical method of Section~\ref{sec:numericalCurlyS} for the case $M=3$, $\alpha = 1$. We used the following choice of cut-off $L$ and number of discretisation points $m$: for $\lambda = 1$: $L=31$, $m=50$; for $\lambda = 10$: $L =27$, $m=50$; for $\lambda = 100$: $L =27$, $m=100$; for $\lambda = 10000$: $L =26$, $m=450$; for $\lambda = 100000$: $L =26$, $m=1350$; for $\lambda = 200000$: $L =26$, $m=1950$. By changing the values of $L$ and $m$, we estimate that this leads to around 12 digits precision. }\label{tab:SevenSodd3p1}
\end{table}

\begin{table}[H]
	\centering
	$$
	\begin{array}{|c|c|c|c|}
		\hline
		\lambda & \mathcal{R}_1 & \mathcal{R}_4 & \mathcal{R}\\
\hline
10000&-2.082727208&-2.077653901&-4.160381109\\
100000&-6.587882595&-6.586294408&-13.174177\\
200000&-9.316811034&-9.315689494&-18.63250053\\
300000&-11.41077309&-11.40985789&-22.82063098\\
400000&-13.17605854&-13.17526622&-26.35132475\\
500000&-14.73130324&-14.73059473&-29.46189797\\
\hline
	\end{array}
	$$
	\caption{Values of $\mathcal{R}_1(\lambda)$, $\mathcal{R}_4(\lambda)$ and $\mathcal{R}(\lambda)$, computed with the numerical method explained in Section~\ref{subsec:NumericalEvaluationR}}\label{tab:calI1and4}
\end{table}


\bibliography{Quiverbib}

\providecommand{\href}[2]{#2}\begingroup\raggedright\begin{thebibliography}{10}

\bibitem{Pestun:2007rz}
V.~Pestun, \emph{{Localization of gauge theory on a four-sphere and supersymmetric Wilson loops}}, \href{http://dx.doi.org/10.1007/s00220-012-1485-0}{\emph{Commun. Math. Phys.} {\bf 313} (2012) 71--129}, [\href{https://arxiv.org/abs/0712.2824}{{\tt 0712.2824}}].

\bibitem{Passerini:2011fe}
F.~Passerini and K.~Zarembo, \emph{{Wilson Loops in N=2 Super-Yang-Mills from Matrix Model}}, \href{http://dx.doi.org/10.1007/JHEP09(2011)102}{\emph{JHEP} {\bf 09} (2011) 102}, [\href{https://arxiv.org/abs/1106.5763}{{\tt 1106.5763}}].

\bibitem{Beccaria:2021vuc}
M.~Beccaria, G.~V. Dunne and A.~A. Tseytlin, \emph{{BPS Wilson loop in $ \mathcal{N} $ = 2 superconformal SU(N) \textquotedblleft{}orientifold\textquotedblright{} gauge theory and weak-strong coupling interpolation}}, \href{http://dx.doi.org/10.1007/JHEP07(2021)085}{\emph{JHEP} {\bf 07} (2021) 085}, [\href{https://arxiv.org/abs/2104.12625}{{\tt 2104.12625}}].

\bibitem{Gerchkovitz:2016gxx}
E.~Gerchkovitz, J.~Gomis, N.~Ishtiaque, A.~Karasik, Z.~Komargodski and S.~S. Pufu, \emph{{Correlation Functions of Coulomb Branch Operators}}, \href{http://dx.doi.org/10.1007/JHEP01(2017)103}{\emph{JHEP} {\bf 01} (2017) 103}, [\href{https://arxiv.org/abs/1602.05971}{{\tt 1602.05971}}].

\bibitem{Baggio:2016skg}
M.~Baggio, V.~Niarchos, K.~Papadodimas and G.~Vos, \emph{{Large-N correlation functions in $ \mathcal{N} $ = 2 superconformal QCD}}, \href{http://dx.doi.org/10.1007/JHEP01(2017)101}{\emph{JHEP} {\bf 01} (2017) 101}, [\href{https://arxiv.org/abs/1610.07612}{{\tt 1610.07612}}].

\bibitem{Rodriguez-Gomez:2016ijh}
D.~Rodriguez-Gomez and J.~G. Russo, \emph{{Large N Correlation Functions in Superconformal Field Theories}}, \href{http://dx.doi.org/10.1007/JHEP06(2016)109}{\emph{JHEP} {\bf 06} (2016) 109}, [\href{https://arxiv.org/abs/1604.07416}{{\tt 1604.07416}}].

\bibitem{Rodriguez-Gomez:2016cem}
D.~Rodriguez-Gomez and J.~G. Russo, \emph{{Operator mixing in large $N$ superconformal field theories on S$^{4}$ and correlators with Wilson loops}}, \href{http://dx.doi.org/10.1007/JHEP12(2016)120}{\emph{JHEP} {\bf 12} (2016) 120}, [\href{https://arxiv.org/abs/1607.07878}{{\tt 1607.07878}}].

\bibitem{Billo:2017glv}
M.~Billo, F.~Fucito, A.~Lerda, J.~F. Morales, Y.~S. Stanev and C.~Wen, \emph{{Two-point correlators in $N =2$ gauge theories}}, \href{http://dx.doi.org/10.1016/j.nuclphysb.2017.11.003}{\emph{Nucl. Phys. B} {\bf 926} (2018) 427--466}, [\href{https://arxiv.org/abs/1705.02909}{{\tt 1705.02909}}].

\bibitem{Beccaria:2020hgy}
M.~Beccaria, M.~Bill\`o, F.~Galvagno, A.~Hasan and A.~Lerda, \emph{{$ \mathcal{N} $ = 2 Conformal SYM theories at large $ \mathcal{N} $}}, \href{http://dx.doi.org/10.1007/JHEP09(2020)116}{\emph{JHEP} {\bf 09} (2020) 116}, [\href{https://arxiv.org/abs/2007.02840}{{\tt 2007.02840}}].

\bibitem{Beccaria:2021hvt}
M.~Beccaria, M.~Bill\`o, M.~Frau, A.~Lerda and A.~Pini, \emph{{Exact results in a $ \mathcal{N} $ = 2 superconformal gauge theory at strong coupling}}, \href{http://dx.doi.org/10.1007/JHEP07(2021)185}{\emph{JHEP} {\bf 07} (2021) 185}, [\href{https://arxiv.org/abs/2105.15113}{{\tt 2105.15113}}].

\bibitem{Bobev:2022grf}
N.~Bobev, P.-J. De~Smet and X.~Zhang, \emph{{The planar limit of the $ \mathcal{N} $ = 2 E-theory: numerical calculations and the large \ensuremath{\lambda} expansion}}, {\emph{JHEP} {\bf 02} (2024) 100}, [\href{https://arxiv.org/abs/2207.12843}{{\tt 2207.12843}}].

\bibitem{Pini:2017ouj}
A.~Pini, D.~Rodriguez-Gomez and J.~G. Russo, \emph{{Large $N$ correlation functions $ \mathcal{N}=$ 2 superconformal quivers}}, \href{http://dx.doi.org/10.1007/JHEP08(2017)066}{\emph{JHEP} {\bf 08} (2017) 066}, [\href{https://arxiv.org/abs/1701.02315}{{\tt 1701.02315}}].

\bibitem{Billo:2018oog}
M.~Billo, F.~Galvagno, P.~Gregori and A.~Lerda, \emph{{Correlators between Wilson loop and chiral operators in $ \mathcal{N}=2 $ conformal gauge theories}}, \href{http://dx.doi.org/10.1007/JHEP03(2018)193}{\emph{JHEP} {\bf 03} (2018) 193}, [\href{https://arxiv.org/abs/1802.09813}{{\tt 1802.09813}}].

\bibitem{Pini:2023svd}
A.~Pini and P.~Vallarino, \emph{{Defect correlators in a $ \mathcal{N} $ = 2 SCFT at strong coupling}}, \href{http://dx.doi.org/10.1007/JHEP06(2023)050}{\emph{JHEP} {\bf 06} (2023) 050}, [\href{https://arxiv.org/abs/2303.08210}{{\tt 2303.08210}}].

\bibitem{Sysoeva:2017fhr}
E.~Sysoeva, \emph{{Wilson loops and its correlators with chiral operators in $\mathcal{N}=2, 4$ SCFT at large $N$}}, \href{http://dx.doi.org/10.1007/JHEP03(2018)155}{\emph{JHEP} {\bf 03} (2018) 155}, [\href{https://arxiv.org/abs/1712.10297}{{\tt 1712.10297}}].

\bibitem{Ennes:2000fu}
I.~P. Ennes, C.~Lozano, S.~G. Naculich and H.~J. Schnitzer, \emph{{Elliptic models, type IIB orientifolds and the AdS / CFT correspondence}}, \href{http://dx.doi.org/10.1016/S0550-3213(00)00580-0}{\emph{Nucl. Phys. B} {\bf 591} (2000) 195--226}, [\href{https://arxiv.org/abs/hep-th/0006140}{{\tt hep-th/0006140}}].

\bibitem{Kachru:1998ys}
S.~Kachru and E.~Silverstein, \emph{{4-D conformal theories and strings on orbifolds}}, \href{http://dx.doi.org/10.1103/PhysRevLett.80.4855}{\emph{Phys. Rev. Lett.} {\bf 80} (1998) 4855--4858}, [\href{https://arxiv.org/abs/hep-th/9802183}{{\tt hep-th/9802183}}].

\bibitem{Gukov:1998kk}
S.~Gukov, \emph{{Comments on N=2 AdS orbifolds}}, \href{http://dx.doi.org/10.1016/S0370-2693(98)01005-3}{\emph{Phys. Lett. B} {\bf 439} (1998) 23--28}, [\href{https://arxiv.org/abs/hep-th/9806180}{{\tt hep-th/9806180}}].

\bibitem{Billo:2021rdb}
M.~Billo, M.~Frau, F.~Galvagno, A.~Lerda and A.~Pini, \emph{{Strong-coupling results for $ \mathcal{N} $ = 2 superconformal quivers and holography}}, \href{http://dx.doi.org/10.1007/JHEP10(2021)161}{\emph{JHEP} {\bf 10} (2021) 161}, [\href{https://arxiv.org/abs/2109.00559}{{\tt 2109.00559}}].

\bibitem{Billo:2022fnb}
M.~Billo, M.~Frau, A.~Lerda, A.~Pini and P.~Vallarino, \emph{{Localization vs holography in 4d$ \mathcal{N} $ = 2 quiver theories}}, \href{http://dx.doi.org/10.1007/JHEP10(2022)020}{\emph{JHEP} {\bf 10} (2022) 020}, [\href{https://arxiv.org/abs/2207.08846}{{\tt 2207.08846}}].

\bibitem{Mitev:2014yba}
V.~Mitev and E.~Pomoni, \emph{{Exact effective couplings of four dimensional gauge theories with $\mathcal N=$ 2 supersymmetry}}, \href{http://dx.doi.org/10.1103/PhysRevD.92.125034}{\emph{Phys. Rev. D} {\bf 92} (2015) 125034}, [\href{https://arxiv.org/abs/1406.3629}{{\tt 1406.3629}}].

\bibitem{Mitev:2015oty}
V.~Mitev and E.~Pomoni, \emph{{Exact Bremsstrahlung and Effective Couplings}}, \href{http://dx.doi.org/10.1007/JHEP06(2016)078}{\emph{JHEP} {\bf 06} (2016) 078}, [\href{https://arxiv.org/abs/1511.02217}{{\tt 1511.02217}}].

\bibitem{Galvagno:2020cgq}
F.~Galvagno and M.~Preti, \emph{{Chiral correlators in $ \mathcal{N} $ = 2 superconformal quivers}}, \href{http://dx.doi.org/10.1007/JHEP05(2021)201}{\emph{JHEP} {\bf 05} (2021) 201}, [\href{https://arxiv.org/abs/2012.15792}{{\tt 2012.15792}}].

\bibitem{Billo:2022gmq}
M.~Bill\`o, M.~Frau, A.~Lerda, A.~Pini and P.~Vallarino, \emph{{Structure Constants in N=2 Superconformal Quiver Theories at Strong Coupling and Holography}}, \href{http://dx.doi.org/10.1103/PhysRevLett.129.031602}{\emph{Phys. Rev. Lett.} {\bf 129} (2022) 031602}, [\href{https://arxiv.org/abs/2206.13582}{{\tt 2206.13582}}].

\bibitem{Beccaria:2022ypy}
M.~Beccaria, G.~P. Korchemsky and A.~A. Tseytlin, \emph{{Strong coupling expansion in \ensuremath{\mathcal{N}} = 2 superconformal theories and the Bessel kernel}}, \href{http://dx.doi.org/10.1007/JHEP09(2022)226}{\emph{JHEP} {\bf 09} (2022) 226}, [\href{https://arxiv.org/abs/2207.11475}{{\tt 2207.11475}}].

\bibitem{Billo:2022lrv}
M.~Billo, M.~Frau, A.~Lerda, A.~Pini and P.~Vallarino, \emph{{Strong coupling expansions in $ \mathcal{N} $ = 2 quiver gauge theories}}, \href{http://dx.doi.org/10.1007/JHEP01(2023)119}{\emph{JHEP} {\bf 01} (2023) 119}, [\href{https://arxiv.org/abs/2211.11795}{{\tt 2211.11795}}].

\bibitem{Rey:2010ry}
S.-J. Rey and T.~Suyama, \emph{{Exact Results and Holography of Wilson Loops in N=2 Superconformal (Quiver) Gauge Theories}}, \href{http://dx.doi.org/10.1007/JHEP01(2011)136}{\emph{JHEP} {\bf 01} (2011) 136}, [\href{https://arxiv.org/abs/1001.0016}{{\tt 1001.0016}}].

\bibitem{Zarembo:2020tpf}
K.~Zarembo, \emph{{Quiver CFT at strong coupling}}, \href{http://dx.doi.org/10.1007/JHEP06(2020)055}{\emph{JHEP} {\bf 06} (2020) 055}, [\href{https://arxiv.org/abs/2003.00993}{{\tt 2003.00993}}].

\bibitem{Ouyang:2020hwd}
H.~Ouyang, \emph{{Wilson loops in circular quiver SCFTs at strong coupling}}, \href{http://dx.doi.org/10.1007/JHEP02(2021)178}{\emph{JHEP} {\bf 02} (2021) 178}, [\href{https://arxiv.org/abs/2011.03531}{{\tt 2011.03531}}].

\bibitem{Fiol:2020ojn}
B.~Fiol, J.~Martfnez-Montoya and A.~Rios~Fukelman, \emph{{The planar limit of $ \mathcal{N} $ = 2 superconformal quiver theories}}, \href{http://dx.doi.org/10.1007/JHEP08(2020)161}{\emph{JHEP} {\bf 08} (2020) 161}, [\href{https://arxiv.org/abs/2006.06379}{{\tt 2006.06379}}].

\bibitem{Galvagno:2021bbj}
F.~Galvagno and M.~Preti, \emph{{Wilson loop correlators in $ \mathcal{N} $ = 2 superconformal quivers}}, \href{http://dx.doi.org/10.1007/JHEP11(2021)023}{\emph{JHEP} {\bf 11} (2021) 023}, [\href{https://arxiv.org/abs/2105.00257}{{\tt 2105.00257}}].

\bibitem{Preti:2022inu}
M.~Preti, \emph{{Correlators in superconformal quivers made QUICK}},  \href{https://arxiv.org/abs/2212.14823}{{\tt 2212.14823}}.

\bibitem{Beccaria:2023kbl}
M.~Beccaria, G.~P. Korchemsky and A.~A. Tseytlin, \emph{{Non-planar corrections in orbifold/orientifold $ \mathcal{N} $ = 2 superconformal theories from localization}}, \href{http://dx.doi.org/10.1007/JHEP05(2023)165}{\emph{JHEP} {\bf 05} (2023) 165}, [\href{https://arxiv.org/abs/2303.16305}{{\tt 2303.16305}}].

\bibitem{Beccaria:2023qnu}
M.~Beccaria and G.~P. Korchemsky, \emph{{Four-dimensional $\mathcal{N}$ = 2 superconformal long circular quivers}}, \href{http://dx.doi.org/10.1007/JHEP04(2024)054}{\emph{JHEP} {\bf 04} (2024) 054}, [\href{https://arxiv.org/abs/2312.03836}{{\tt 2312.03836}}].

\bibitem{Sobko:2025zci}
E.~Sobko, \emph{{Continuous quiver gauge theories}}, \href{http://dx.doi.org/10.1103/PhysRevD.111.046022}{\emph{Phys. Rev. D} {\bf 111} (2025) 046022}.

\bibitem{Pini:2024uia}
A.~Pini and P.~Vallarino, \emph{{Integrated correlators at strong coupling in an orbifold of $ \mathcal{N} $ = 4 SYM}}, \href{http://dx.doi.org/10.1007/JHEP06(2024)170}{\emph{JHEP} {\bf 06} (2024) 170}, [\href{https://arxiv.org/abs/2404.03466}{{\tt 2404.03466}}].

\bibitem{Bajnok:2024epf}
Z.~Bajnok, B.~Boldis and G.~P. Korchemsky, \emph{{Tracy-Widom Distribution in Four-Dimensional Supersymmetric Yang-Mills Theories}}, \href{http://dx.doi.org/10.1103/PhysRevLett.133.031601}{\emph{Phys. Rev. Lett.} {\bf 133} (2024) 031601}, [\href{https://arxiv.org/abs/2403.13050}{{\tt 2403.13050}}].

\bibitem{Bajnok:2024ymr}
Z.~Bajnok, B.~Boldis and G.~P. Korchemsky, \emph{{Solving four-dimensional superconformal Yang-Mills theories with Tracy-Widom distribution}},  \href{https://arxiv.org/abs/2409.17227}{{\tt 2409.17227}}.

\bibitem{Skrzypek:2023fkr}
T.~Skrzypek and A.~A. Tseytlin, \emph{{On AdS/CFT duality in the twisted sector of string theory on AdS$_{5}$\texttimes{} S$^{5}$/\ensuremath{\mathbb{Z}}$_{2}$ orbifold background}}, \href{http://dx.doi.org/10.1007/JHEP03(2024)045}{\emph{JHEP} {\bf 03} (2024) 045}, [\href{https://arxiv.org/abs/2312.13850}{{\tt 2312.13850}}].

\bibitem{Korchemsky:2025eyc}
G.~P. Korchemsky and A.~Testa, \emph{{Correlation functions in four-dimensional superconformal long circular quivers}},  \href{https://arxiv.org/abs/2501.17223}{{\tt 2501.17223}}.

\bibitem{Ferrando:2025qkr}
G.~Ferrando, S.~Komatsu, G.~Lefundes and D.~Serban, \emph{{Exact Three-Point Functions in $\mathcal{N}=2$ Superconformal Field Theories: Integrability vs. Localization}},  \href{https://arxiv.org/abs/2503.07295}{{\tt 2503.07295}}.

\bibitem{Belitsky:2020qrm}
A.~V. Belitsky and G.~P. Korchemsky, \emph{{Octagon at finite coupling}}, \href{http://dx.doi.org/10.1007/JHEP07(2020)219}{\emph{JHEP} {\bf 07} (2020) 219}, [\href{https://arxiv.org/abs/2003.01121}{{\tt 2003.01121}}].

\bibitem{Belitsky:2020qir}
A.~V. Belitsky and G.~P. Korchemsky, \emph{{Crossing bridges with strong Szeg\H{o} limit theorem}}, \href{http://dx.doi.org/10.1007/JHEP04(2021)257}{\emph{JHEP} {\bf 04} (2021) 257}, [\href{https://arxiv.org/abs/2006.01831}{{\tt 2006.01831}}].

\bibitem{Pini:2023lyo}
A.~Pini and P.~Vallarino, \emph{{Wilson loop correlators at strong coupling in $ \mathcal{N} $ = 2 quiver gauge theories}}, \href{http://dx.doi.org/10.1007/JHEP11(2023)003}{\emph{JHEP} {\bf 11} (2023) 003}, [\href{https://arxiv.org/abs/2308.03848}{{\tt 2308.03848}}].

\bibitem{Costin:2020hwg}
O.~Costin and G.~V. Dunne, \emph{{Physical Resurgent Extrapolation}}, \href{http://dx.doi.org/10.1016/j.physletb.2020.135627}{\emph{Phys. Lett. B} {\bf 808} (2020) 135627}, [\href{https://arxiv.org/abs/2003.07451}{{\tt 2003.07451}}].

\bibitem{Costin:2020pcj}
O.~Costin and G.~V. Dunne, \emph{{Uniformization and Constructive Analytic Continuation of Taylor Series}}, \href{http://dx.doi.org/10.1007/s00220-022-04361-6}{\emph{Commun. Math. Phys.} {\bf 392} (2022) 863--906}, [\href{https://arxiv.org/abs/2009.01962}{{\tt 2009.01962}}].

\bibitem{Pufu:2023vwo}
S.~S. Pufu, V.~A. Rodriguez and Y.~Wang, \emph{{Scattering From $(p,q)$-Strings in $\text{AdS}_5 \times \text{S}^5$}},  \href{https://arxiv.org/abs/2305.08297}{{\tt 2305.08297}}.

\bibitem{Billo:2023ncz}
M.~Billo', F.~Galvagno, M.~Frau and A.~Lerda, \emph{{Integrated correlators with a Wilson line in $ \mathcal{N} $ = 4 SYM}}, \href{http://dx.doi.org/10.1007/JHEP12(2023)047}{\emph{JHEP} {\bf 12} (2023) 047}, [\href{https://arxiv.org/abs/2308.16575}{{\tt 2308.16575}}].

\bibitem{Douglas:1996sw}
M.~R. Douglas and G.~W. Moore, \emph{{D-branes, quivers, and ALE instantons}},  \href{https://arxiv.org/abs/hep-th/9603167}{{\tt hep-th/9603167}}.

\bibitem{Maldacena:1998im}
J.~M. Maldacena, \emph{{Wilson loops in large N field theories}}, \href{http://dx.doi.org/10.1103/PhysRevLett.80.4859}{\emph{Phys. Rev. Lett.} {\bf 80} (1998) 4859--4862}, [\href{https://arxiv.org/abs/hep-th/9803002}{{\tt hep-th/9803002}}].

\bibitem{Semenoff:2001xp}
G.~W. Semenoff and K.~Zarembo, \emph{{More exact predictions of SUSYM for string theory}}, \href{http://dx.doi.org/10.1016/S0550-3213(01)00455-2}{\emph{Nucl. Phys. B} {\bf 616} (2001) 34--46}, [\href{https://arxiv.org/abs/hep-th/0106015}{{\tt hep-th/0106015}}].

\bibitem{Dempsey:2024vkf}
R.~Dempsey, B.~Offertaler, S.~S. Pufu and Y.~Wang, \emph{{Global Symmetry and Integral Constraint on Superconformal Lines in Four Dimensions}},  \href{https://arxiv.org/abs/2405.10914}{{\tt 2405.10914}}.

\bibitem{Billo:2024kri}
M.~Bill\`o, M.~Frau, F.~Galvagno and A.~Lerda, \emph{{A note on integrated correlators with a Wilson line in $\mathcal{N}=4$ SYM}},  \href{https://arxiv.org/abs/2405.10862}{{\tt 2405.10862}}.

\bibitem{Dorigoni:2024vrb}
D.~Dorigoni, Z.~Duan, D.~R. Pavarini, C.~Wen and H.~Xie, \emph{{Electromagnetic duality for line defect correlators in $ \mathcal{N} $ = 4 super Yang-Mills theory}}, \href{http://dx.doi.org/10.1007/JHEP11(2024)084}{\emph{JHEP} {\bf 11} (2024) 084}, [\href{https://arxiv.org/abs/2409.12786}{{\tt 2409.12786}}].

\bibitem{Dorigoni:2024csx}
D.~Dorigoni, \emph{{Note on \textquoteright{}t Hooft-line defect integrated correlators in N=4 supersymmetric Yang-Mills theory}}, \href{http://dx.doi.org/10.1103/PhysRevD.110.L121702}{\emph{Phys. Rev. D} {\bf 110} (2024) L121702}, [\href{https://arxiv.org/abs/2410.02377}{{\tt 2410.02377}}].

\bibitem{Pini:2024zwi}
A.~Pini, \emph{{Integrated correlators with a Wilson line in a $ \mathcal{N} $ = 2 quiver gauge theory at strong coupling}}, \href{http://dx.doi.org/10.1007/JHEP01(2025)195}{\emph{JHEP} {\bf 01} (2025) 195}, [\href{https://arxiv.org/abs/2410.17342}{{\tt 2410.17342}}].

\bibitem{DeLillo:2025hal}
L.~De~Lillo, M.~Frau and A.~Pini, \emph{{Integrated line-defect correlators in $Sp(N)$ SCFTs at strong coupling}},  \href{https://arxiv.org/abs/2503.04902}{{\tt 2503.04902}}.

\bibitem{Chester:2022sqb}
S.~M. Chester, \emph{{Bootstrapping 4d $ \mathcal{N} $ = 2 gauge theories: the case of SQCD}}, \href{http://dx.doi.org/10.1007/JHEP01(2023)107}{\emph{JHEP} {\bf 01} (2023) 107}, [\href{https://arxiv.org/abs/2205.12978}{{\tt 2205.12978}}].

\bibitem{Alday:2024srr}
L.~F. Alday and X.~Zhou, \emph{{Flat-space limit of defect correlators and stringy AdS form factors}}, \href{http://dx.doi.org/10.1007/JHEP03(2025)182}{\emph{JHEP} {\bf 03} (2025) 182}, [\href{https://arxiv.org/abs/2411.04378}{{\tt 2411.04378}}].

\bibitem{Fiol:2018yuc}
B.~Fiol, J.~Mart\'\i{}nez-Montoya and A.~Rios~Fukelman, \emph{{Wilson loops in terms of color invariants}}, \href{http://dx.doi.org/10.1007/JHEP05(2019)202}{\emph{JHEP} {\bf 05} (2019) 202}, [\href{https://arxiv.org/abs/1812.06890}{{\tt 1812.06890}}].

\bibitem{Erickson:2000af}
J.~K. Erickson, G.~W. Semenoff and K.~Zarembo, \emph{{Wilson loops in N=4 supersymmetric Yang-Mills theory}}, \href{http://dx.doi.org/10.1016/S0550-3213(00)00300-X}{\emph{Nucl. Phys. B} {\bf 582} (2000) 155--175}, [\href{https://arxiv.org/abs/hep-th/0003055}{{\tt hep-th/0003055}}].

\bibitem{waldvogel2006fast}
J.~Waldvogel, \emph{Fast construction of the {Fej{\'e}r} and {Clenshaw}--{Curtis} quadrature rules}, {\emph{BIT Numerical Mathematics} {\bf 46} (2006) 195--202}.

\bibitem{WolframFunctionRepository}
``{FejerQuadratureWeights}.'' \url{https://resources.wolframcloud.com/FunctionRepository/resources/FejerQuadratureWeights/}.

\bibitem{Billo:2022xas}
M.~Billo, M.~Frau, A.~Lerda, A.~Pini and P.~Vallarino, \emph{{Three-point functions in a $ \mathcal{N} $ = 2 superconformal gauge theory and their strong-coupling limit}}, \href{http://dx.doi.org/10.1007/JHEP08(2022)199}{\emph{JHEP} {\bf 08} (2022) 199}, [\href{https://arxiv.org/abs/2202.06990}{{\tt 2202.06990}}].

\bibitem{Beccaria:2021ism}
M.~Beccaria, G.~V. Dunne and A.~A. Tseytlin, \emph{{Strong coupling expansion of free energy and BPS Wilson loop in $ \mathcal{N} $ = 2 superconformal models with fundamental hypermultiplets}}, \href{http://dx.doi.org/10.1007/JHEP08(2021)102}{\emph{JHEP} {\bf 08} (2021) 102}, [\href{https://arxiv.org/abs/2105.14729}{{\tt 2105.14729}}].

\bibitem{Billo:2024ftq}
M.~Billo, M.~Frau, A.~Lerda, A.~Pini and P.~Vallarino, \emph{{Integrated correlators in a $ \mathcal{N} $ = 2 SYM theory with fundamental flavors: a matrix-model perspective}}, \href{http://dx.doi.org/10.1007/JHEP11(2024)172}{\emph{JHEP} {\bf 11} (2024) 172}, [\href{https://arxiv.org/abs/2407.03509}{{\tt 2407.03509}}].

\bibitem{Binder:2019jwn}
D.~J. Binder, S.~M. Chester, S.~S. Pufu and Y.~Wang, \emph{{$ \mathcal{N} $ = 4 Super-Yang-Mills correlators at strong coupling from string theory and localization}}, \href{http://dx.doi.org/10.1007/JHEP12(2019)119}{\emph{JHEP} {\bf 12} (2019) 119}, [\href{https://arxiv.org/abs/1902.06263}{{\tt 1902.06263}}].

\bibitem{Billo:2023kak}
M.~Billo, M.~Frau, A.~Lerda and A.~Pini, \emph{{A matrix-model approach to integrated correlators in a $ \mathcal{N} $ = 2 SYM theory}}, \href{http://dx.doi.org/10.1007/JHEP01(2024)154}{\emph{JHEP} {\bf 01} (2024) 154}, [\href{https://arxiv.org/abs/2311.17178}{{\tt 2311.17178}}].

\end{thebibliography}\endgroup
\bibliographystyle{JHEP}

\end{document}